\documentclass[twocolumn]{aastex631}

\newcommand{\Msun}{\,\ensuremath{\mathrm{M}_\odot}}

\shorttitle{He and Sr in kilonova}
\shortauthors{Tarumi et al.}
\graphicspath{{./}{figures/}}

\begin{document}

\title{Non-LTE analysis for Helium and Strontium lines in the kilonova AT2017gfo}

\correspondingauthor{Yuta Tarumi}
\email{yuta.tarumi@phys.s.u-tokyo.ac.jp}
\author{Yuta Tarumi}
\affiliation{Department of Physics, School of Science, The University of Tokyo, Bunkyo, Tokyo 113-0033, Japan}

\author{Kenta Hotokezaka}
\affiliation{Department of Physics, School of Science, The University of Tokyo, Bunkyo, Tokyo 113-0033, Japan}

\author{Nanae Domoto}
\affiliation{Astronomical Institute, Tohoku University, Aoba, Sendai 980-8578, Japan}

\author{Masaomi Tanaka}
\affiliation{Astronomical Institute, Tohoku University, Aoba, Sendai 980-8578, Japan}
\affiliation{Division for the Establishment of Frontier Sciences, Organization for Advanced Studies, Tohoku University, Aoba, Sendai 980-8577, Japan}



\begin{abstract}

Kilonova spectra imprint valuable information about the elements synthesized in neutron star mergers. In AT2017gfo, the kilonova associated with GW170817, the spectroscopic feature centered around 8000 angstroms has been interpreted as the P-Cygni profile arising from singly ionized Strontium. Recently, \citet{2022Perego} suggested that Helium 10833 line can be an alternative explanation of the feature. Here, we study the line features under non-local thermodynamic equilibrium. We find that the ionization of ejecta by the stopping of radioactive decay product can significantly enhance the ionization states around the line forming region. We compute the kilonova spectrum under the assumption of spherical symmetry and uniform elemental fraction in the line-forming region. We find that 0.2\% (in mass) of Helium in the ejecta can reproduce the P-Cygni feature in the observed spectrum at $1.43$ -- $4.40$ days. Strontium with a mass fraction of $1\%$ is also able to make the absorption feature at $\sim 1.5\,$days, but it gets weaker with time due to ionization by radioactive decay products. 
The strength of the He line signature depends sensitively on UV strength for the first two epochs. Further modeling of UV line blanketing by $r$-process elements and the optical properties of light $r$-process elements would be crucial to distinguish between Helium and Strontium features. The mass fraction of He is a good indicator for ejecta entropy that allows us to probe the mass ejection mechanism.

\end{abstract}



\section{Introduction} \label{sec:intro}
Radioactive decays of neutron-rich nuclei synthesized  in neutron star merger ejecta through rapid capture of neutrons (``{\it r}-process'')  power an electromagnetic transient, ``kilonova'' (kN, \citealt{Metzger2010MNRAS}). Currently, only one kN (AT2017gfo) that is associated with a gravitational-wave event (GW170817) has been observed \citep[see][for  reviews]{Metzger2017LRR,Nakar2020PhR,Margutti2021ARA&A}. The light curve and spectrum of kNe imprint valuable information about the elements synthesized in the merger ejecta \citep{Barnes2013ApJ,Kasen2013ApJ,Kasen2017,Tanaka2013ApJ,Tanaka2018ApJ,Kawaguchi2018ApJ,Wollaeger2018MNRAS,2018Wanajo,Waxman2019ApJ,WuPhRvL,Banerjee2020ApJ,Hotokezaka2020ApJ,Barnes2021ApJ,Korobkin2021ApJ,Domoto2021,Domoto2022,Gillanders2022,2022Perego}. 

The spectra of the kN AT2017gfo in the optical and near-infrared bands were obtained by various groups \citep{Chornock2017ApJ,Kasliwal2017,Kilpatrick2017,Pian2017,Smartt2017}. One of the most remarkable features in the spectra is a strong absorption observed around $8000{\rm \AA}$ existing from 1.5 days to about a week after the merger \citep{Kilpatrick2017,Pian2017}.
Given the fact that the ejecta is expected to be composed of many heavy elements, such  absorption features in the spectra may arise as a result of several absorption lines of multiple ion species. However,  because the number density of strong lines decreases at longer wavelengths, e.g., $\gtrsim 10000{\rm \AA}$ \citep{Domoto2022}, it is worth investigating whether a single ion species can produce observable absorption features.
For instance, \citet{2019Watson} interpret the  $8000{\rm \AA}$ absorption as a P-Cygni feature caused by the triplet lines of  singly ionized Strontium (Sr\,II) at $10039,\,10330,\,10918{\rm \AA}$
blueshifted by $0.2c$ (see also \citealt{Domoto2021,Domoto2022,Gillanders2022}).  If true, this spectral feature is the first direct confirmation of {\it r}-process element synthesis in a neutron star merger.

{There is an alternative explanation of this feature by neutral Helium (He\,I) $10833{\rm \AA}$ absorption line. The presence of He in a significant amount in neutron star merger ejecta is first predicted by \cite{Fernandez2013}. They point out that $\alpha$ particles can remain in the ejecta because of $\alpha$-rich freeze-out depending on the outflow's entropy and electron fraction.
More recently, \citet{2022Perego} show that
$\alpha$-decay of heavy nuclei also contributes to the $\alpha$ particle production and point out that
neutral He can cause a spectral feature similar to the Sr\,II lines.  
However, they have concluded that the feature observed in AT2017gfo is unlikely to be He because luminosity needs to be lower than the observation by order of magnitude with their assumed He mass ($\sim$ a few $10^{-6}\,\Msun$). However, there is a wide range of predictions from simulations on the fraction of He in neutron star merger ejecta \citep{2018Wanajo, 2022Fujibayashi}.}

Most works on  kN radiation transfer in the literature, including \cite{2019Watson}, assume local thermal equilibrium (LTE), where the ionization states and populations of excited levels are determined by assuming Saha equilibrium and Boltzmann distribution of the local temperature. The LTE approximation may be
valid around the typical density in the photospheric phase \citep{Pognan2022MNRAS}. 
However, radioactive ionization can be more important than thermal ionization at lower densities \citep{Hotokezaka2021MNRAS}. 
Furthermore, the first excited level of He\,I $2^3$S, responsible for the $10833{\rm \AA}$ absorption line, is populated only through non-thermal processes at the kN 
photospheric temperatures $\lesssim 5000\,{\rm K}$. Indeed, non-thermal particles play a crucial role in the He line signature in type-Ib supernovae \citep{1991Lucy, 2012Hachinger}.
Therefore, including non-thermal processes may play a significant role in interpreting the spectral features in kNe.

In this paper, we model the level population of He\,I and Sr\,II in kNe {under the existence of non-thermal electrons produced by the $\beta$-decay of $r$-process elements. We will show that  the P-Cygni feature can be reproduced by the $10833{\rm \AA}$ He line with a reasonable He mass. We will also show that, unlike type-Ib supernovae, this is the only absorption line of He observable in the kN photospheric phase. In the case of Sr, the Sr\,II abundance decreases with time because 
ionization of Sr\,II $\rightarrow$ Sr\,III proceeds due to a decline in the recombination rate.
Consequently, the line strength gets weaker from 1.5 days to 4.5 days after the merger, which seems to contradict the observed feature.
} First, we present the picture and summarize the physical processes for modeling the population of excited states in kN in Section 2. In Section 3, we will show the model spectrum and compare them to observations. Section 4 will be used to discuss the implications and limitations of this work.

\section{Basic picture}
\label{sec:basic picture}
We consider a situation where the observed P-Cygni-like feature around $8000{\rm \AA}$ is produced by 
strong lines of He\,I or Sr\,II, of which the elemental fraction is assumed to be spherical and uniform in the line forming region for simplicity. The underlying continuum  is assumed to be a blackbody emission from the spherical photosphere. 
It is also assumed that any scattering and absorption processes other than the He\,I and Sr\,II lines do not play roles in the line-forming region. We briefly introduce the key concepts of our modeling in the following.

\subsection{He and Sr level structures}

\begin{figure}
    \centering
    \includegraphics[width=\columnwidth]{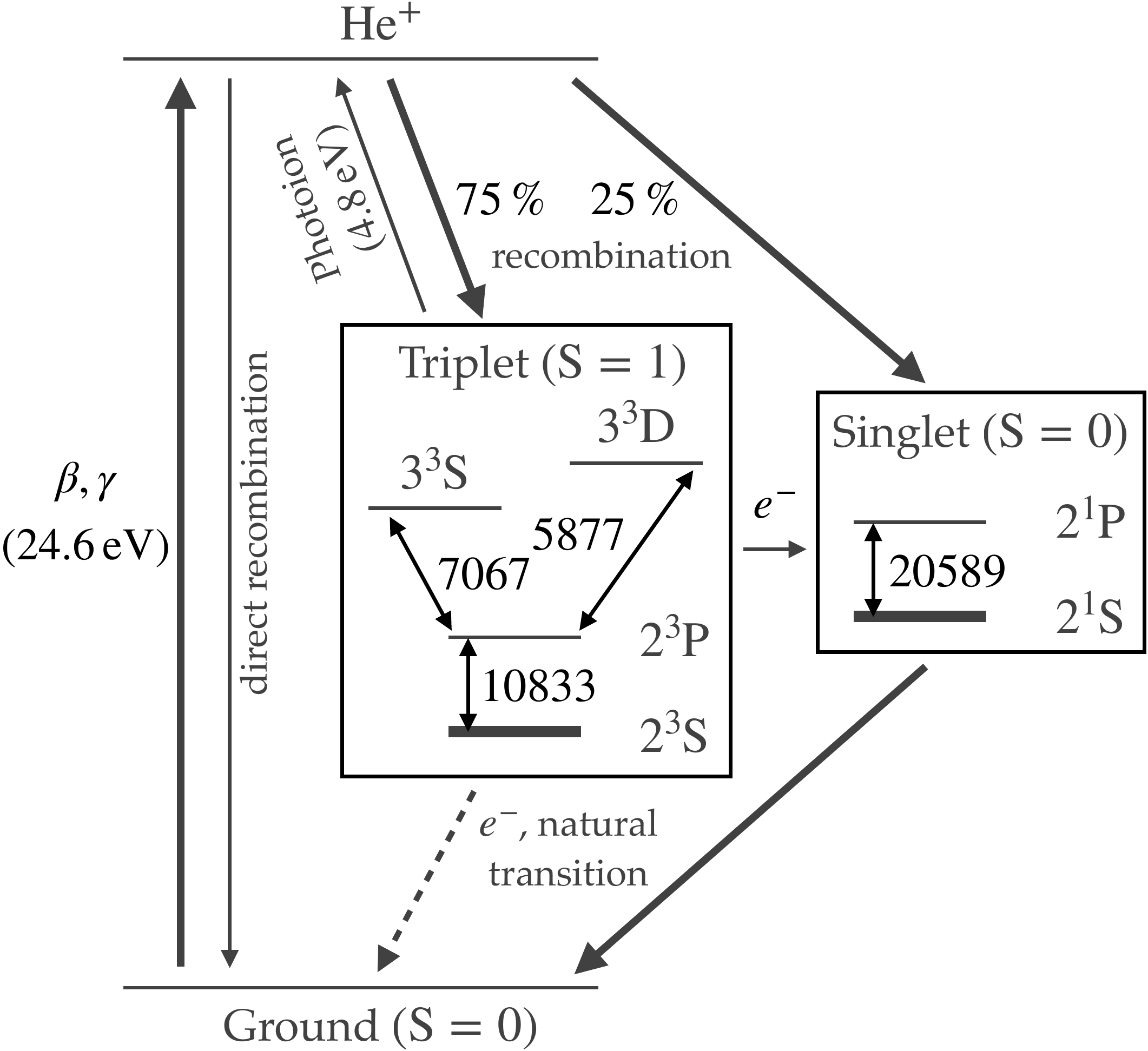}
    \caption{Schematic diagram that shows the populating mechanism of excited He.}
    \label{fig:schematic}
\end{figure}

\begin{figure}
    \centering
    \includegraphics[width=\columnwidth]{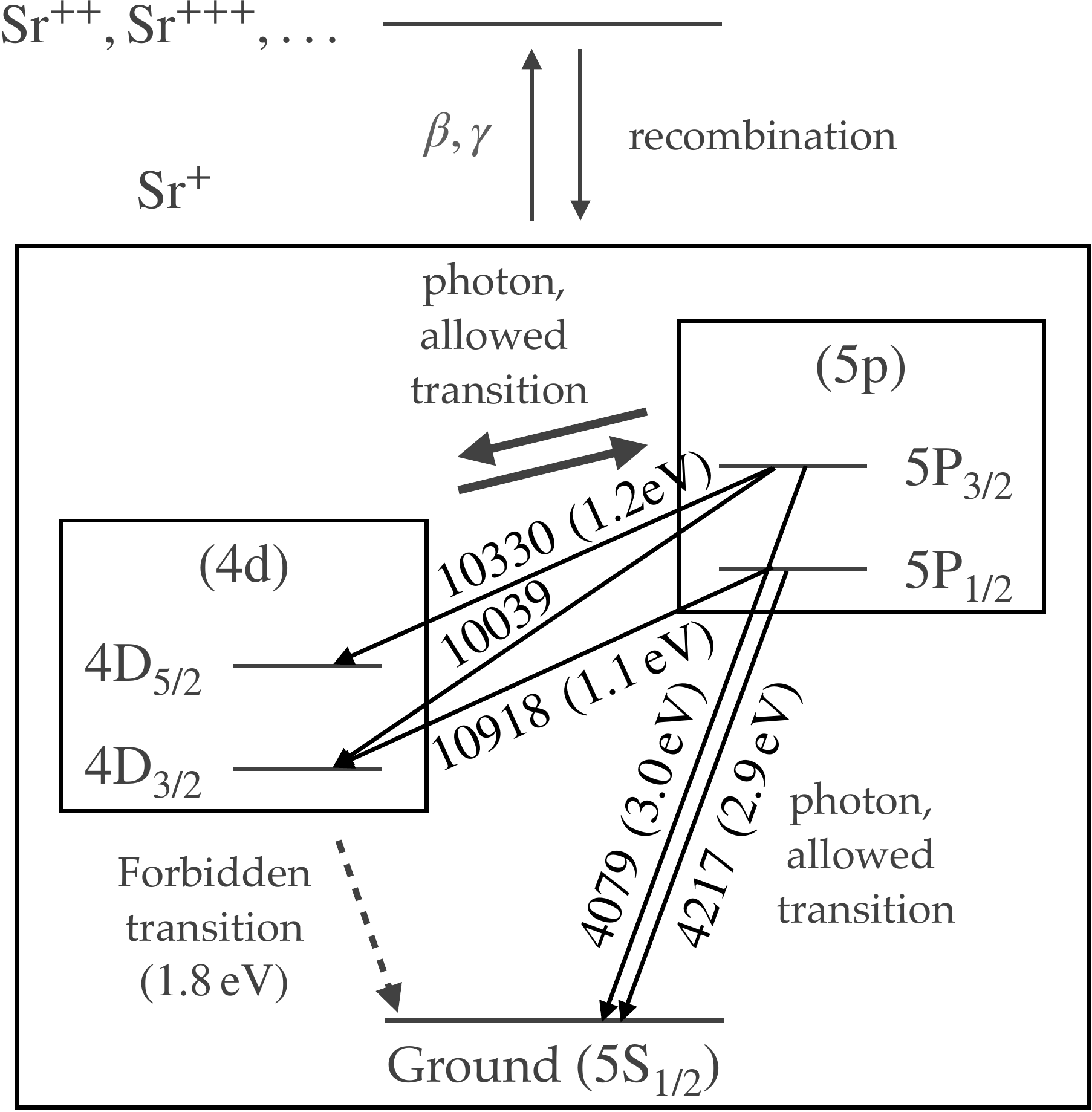}
    \caption{Schematic diagram that shows the populating mechanism of singly ionized Sr.}
    \label{fig:schematic_Sr}
\end{figure}

Here, we describe the level structure and transitions relevant to the He\,I $10833{\rm \AA}$ and Sr\,II triplet $10039,\,10330,\,10918{\rm \AA}$. The He\,I $10833{\rm \AA}$ line arises from transitions between the excited levels $2^3{\rm S}$ and $2^3{\rm P}$, where $2^3{\rm S}$ is a metastable level.
Figure~\ref{fig:schematic} shows a schematic diagram depicting the populating mechanism of the $2^{3}{\rm S}$ level. Note that the excitation energy from the ground level to ${\rm He\,I,2^3{\rm S}}$
is $19.8\,{\rm eV}$, much larger than the thermal energy of kN photospheric temperature at $\lesssim 5000\,{\rm K}$. Although the excitation by any thermal processes is negligible,
${\rm He\,I,2^3{\rm S}}$ is populated through the recombination from He\,II as follows.

Non-thermal particles ($\beta, \gamma$) first ionize neutral He atoms. Singly ionized He captures an electron, and neutral He forms. A good fraction of He\,I are in excited levels just after the capture, and $\sim 75\%$ of excited He\,I will be ``spin-triplet states'' with the total electron spin ${\rm S}=1$ \citep{2006OsterbrockFerland}. He\,I atoms in triplet states decay via natural transition to the lowest-energy triplet states, $2^{3}{\rm S}$. The natural transition to the ground level is highly inefficient, i.e., the lifetime of $\sim 2\,{\rm hr}$, because spin-flip is strongly prohibited. Therefore, the $2^{3}{\rm S}$ level is highly populated if He\,II is efficiently produced through ionization by non-thermal particles.

The triplet lines of Sr\,II arise from transitions between 4D and 5P. The natural decay from 4D to the ground levels is forbidden in electric dipole transitions. However, in an environment with sufficient density of $\sim$ eV thermal photons, the 5P level works as a bypass of the transition, leading the level population to a quasi-thermal distribution. 
Although the effect of non-thermal electrons is not important in excitation they cause the overionization of Sr and reduce the abundance of Sr II, which may significantly reduce the line strength of Sr II in the kN spectra.

\subsection{Line optical depth}

The strength of an absorption line due to an atomic transition $l\rightarrow u$ in an expanding medium may be
estimated by the Sobolev optical depth \citep{1960Sobolev}:
\begin{eqnarray}
        \tau_s\approx 0.23 t_d \lambda_{\rm \mu m} n_l f_{lu},
        \label{eq:Sobolev}
\end{eqnarray}
where $t_d$ is the time since the explosion in units of day,
$\lambda_{\rm \mu m}$ is the line wavelength in units of micron,
$n_l$ is the typical number of the absorbing ions  in the level $l$ per ${\rm cm^{3}}$
in the line-forming region, and $f_{lu}$ is the oscillator strength. 
A line feature may appear in the observed spectra when $\tau_s\gtrsim 1$. Spectra of AT2017gfo show that $\tau_s \simeq 1$ is required to explain the observation (see Section~\ref{sec:optical depth from spectrum}).

For the He\,I lines ($2^3{\rm S}\rightarrow 2^3{\rm P}$) and Sr\,II lines ($4^2{\rm D}\rightarrow 5^2{\rm P}$), the oscillator strengths are $f_{lu}\approx 0.54$ and $0.089$, 
respectively. 
Thus, the conditions of line formation are
\begin{eqnarray}
        n({\rm He\,I,2^3{\rm S}})\gtrsim 7.4t_{d}^{-1}\,{\rm cm^{-3}},\label{eq:HeI_crit} 
\end{eqnarray}

and
\begin{eqnarray}
        n({\rm Sr\,II,4^2{\rm D}})\gtrsim 50t_{d}^{-1}\,{\rm cm^{-3}}.\label{eq:SrII_crit} 
\end{eqnarray}
These critical densities should be compared to the number density of the line-forming levels $n_l$ at a mass shell with an expansion velocity $v$:
\begin{eqnarray}
    n_l(t,v) \approx  10^{6}t_{d}^{-3}\frac{n_{l}}{n_{\rm all}} \frac{m_{\rm ej}(\geq v)}{0.01M_{\odot}}\frac{Y}{0.01}\left(\frac{v}{0.2c} \right)^{-3}\,{\rm cm^{-3}},
\end{eqnarray}
where $Y$ is the number fraction of He and Sr,  $m_{\rm ej}(\geq v)$ is the mass of ejecta faster than $v$. Here we have assumed that the  density $\rho(t,v)$ is $\rho(t,v) \simeq m_{\rm ej}(\geq v)/(vt)^3$ and the mean atomic mass number of ejecta is $100$. He\,I and Sr\,II lines can appear if the fractions of level populations $n_l/n_{\rm all}$ of ${\rm He\,I_{2^3S}}$ and ${\rm Sr\,II_{4^2D}}$ are sufficiently large.

\subsection{Level population under radioactive ionization}
Here, we estimate the populations for the $2^{3}$S level of He\,I and $4^2$D level of Sr\,II and compare them to the critical densities.
The balance between ionization and recombination gives an estimate of the ionization states:
\begin{eqnarray}
    n_{i+1} \approx \frac{\beta_i}{\alpha_in_e}n_i,
\end{eqnarray}
where $\alpha_i$ is the recombination rate coefficient for $i+1\rightarrow i$, $\beta_i$
is the ionization per unit time per ion  for $i\rightarrow i+1$, and $n_e$ is the number density of free electrons. 
The recombination rate coefficients for He are obtained from \citep{2010Nahar}. For Sr, we have scaled the recombination rate coefficient from Hydrogen values with the formula suggested in \citet{1962Bates}: $\alpha(z, T)=z\alpha(1, T/z^2)$, where we take $z$ as the charge of an ion.
$\beta_i$ should include all the relevant ionization processes: thermal ionization, radioactive ionization, and re-ionization by recombination photons.
Here, for simplicity, let us assume that radioactive ionization of $r$-process elements dominates over the other processes:
\begin{eqnarray}
    \beta_{i}(t) = \frac{\dot{q}_{\beta}(t)}{w_i},
    \label{eq:ionization}
\end{eqnarray}
where $\dot{q}_{\beta}\approx 1t_{d}^{-1.3}\,{\rm eV/s}$ \citep{Hotokezaka2020ApJ} is the radioactive heat per unit time per ion, and $w_{i}$ is the required energy to ionize a given ion $i$ (``work per ion pair'').

In astrophysical plasma, the work per ion pair $w_i$ is $10$ -- $100$ times the ionization potential because a few \% of the energy of non-thermal particles is spent to ionize elements. \citet{1980Axelrod} estimates the $w_i$ for iron in type-Ia supernova as $\sim 30$ times the ionization potential. \citet{Hotokezaka2021MNRAS} give similar values for Nd in kilonova ejecta.
We have solved the Spencer-Fano equation to obtain $w_i$ \citep{1954SpencerFano, Kozma1992ApJ}. The calculation details will be presented in our companion paper (Hotokezaka et al. in prep.). 
The values of $w_i$ in neutron star merger ejecta are $\approx 3000\,{\rm eV}$ for He II, $\approx 600\,{\rm eV}$ for He I, $\approx 300\,{\rm eV}$ for Sr II, and $\approx 450\,{\rm eV}$ for Sr III, respectively.

With $\alpha_i$ and $\beta_i$, we then estimate the population ratio of ionization states. Assuming $n_e = 1.5\times 10^8\,t_d^{-3}\,[{\rm cm^{-3}}]$, we estimate
\begin{eqnarray}
    \frac{n({\rm He\,II})}{n({\rm He\,I})} \approx 12 t_{d}^{1.7},~~~\frac{n({\rm He\,III})}{n({\rm He\,II})} \approx 0.5 t_{d}^{1.7}. 
    \label{eq:He ionization}
\end{eqnarray}
At $t_d > 1$, equation~\ref{eq:He ionization} gives the fraction of He\,II as
\begin{eqnarray}
    \biggl(\frac{n_{\rm He\,II}}{n_{\rm He}}\biggr)\approx \left(1+0.5 t_d^{1.7}\right)^{-1}.
\end{eqnarray}

The population of He\,I $2^3$S can be estimated by the balance between the collisional deexcitation and recombination from He\,II:
\begin{eqnarray}
    n({\rm He\,I\,2^3{\rm S}}) & \approx  &\frac{0.75\alpha_{\rm He\,I}}{k}n({\rm He\,II})\nonumber \\
    &\sim & 50\, t_d^{-3}\left(1+0.5t_d^{1.7}\right)^{-1} \nonumber\\
    \times & &\frac{m_{\rm ej}(\geq v )}{0.01M_{\odot}}\frac{Y}{0.01}\left(\frac{v}{0.2c} \right)^{-3}\,{\rm cm^{-3}},
    \label{eq:23S}
\end{eqnarray}
where $\alpha$ is the total recombination rate coefficient, and $k$ is the total collisional depopulation rate of $2^3{\rm S}$. For the details of $k$, see Section~\ref{sec:atomic}.

In the case of Sr, neutral Sr is negligible even in LTE. Assuming the same temperature as He,
\begin{eqnarray}
    \hspace{-5mm}
    \frac{n({\rm Sr\,III})}{n({\rm Sr\,II})} \approx 5t_{d}^{1.7},~~~\frac{n({\rm Sr\,IV})}{n({\rm Sr\,III})} \approx t_{d}^{1.7}.
\end{eqnarray}

Therefore, Sr are mostly Sr\,IV at a few days and
\begin{eqnarray}
    \biggl(\frac{n_{\rm Sr\,II}}{n_{\rm Sr}}\biggr)\approx t_d^{-3.4}/5,
\end{eqnarray}

\begin{eqnarray}
    n({\rm Sr\,II\,4^2D})&\approx &  200 t_{d}^{-6.4}\nonumber\\
    \times & & \frac{m_{\rm ej}(\geq v )}{0.01M_{\odot}}\frac{Y}{0.01}\left(\frac{v}{0.2c} \right)^{-3}\,{\rm cm^{-3}}.
    \label{eq:Sr population}
\end{eqnarray}
Here, we have assumed that the fraction of $4^2$D to the ground level of Sr\,II follows the thermal distribution at $T\sim 3000\,{\rm K}$, which will be justified later. 

The estimates above should be compared to the critical densities for He and Sr (equations~\ref{eq:HeI_crit} and \ref{eq:SrII_crit}). The Sr line signature gets weaker quickly compared to the He line signature (compare equations~\ref{eq:23S} and \ref{eq:Sr population}). We
model the spectral feature of these elements for the kilonova AT2017gfo, taking the time evolution of the photosphere into account for the first few days.

\section{Method}

We model the level population of He and Sr to compute the emergent spectrum. To this end, the rate equations of transitions between various levels are solved by accounting for ionization/excitation by non-thermal particles, spontaneous/stimulated emissions, and electron collisions. In the following, we describe our model to compute the emergent spectrum.

\subsection{Geometry}
\label{sec:geometry}
The ejecta is assumed to be spherically symmetric and homologously expanding with a density profile: 
\begin{equation}
\rho(v, t) = \left\{
\begin{array}{ll}
\rho_0 (v/v_0)^{-p} (t/t_0)^{-3} & (0.1 < v/c < 0.5)\\
0 & ({\rm otherwise})
\end{array}
\right.
\end{equation}
where $\rho_0$ is taken to be consistent with $M_{\rm ej}=0.04 \Msun$ (e.g., \citealt{Smartt2017}). The masses of He and Sr are assumed to be $M_{\rm He}/M_{\rm ej} = 0.2\%$ and $M_{\rm Sr}/M_{\rm ej} = 1\%$, and the composition is homogeneous. We take $p=5$ for our fiducial calculation. 

For simplicity, we assume that the ejecta has a sharply defined photosphere at a photospheric velocity of $v_{\rm phot}$ at each given time. The photosphere is assumed to have a single temperature $T_{\gamma}$, and it emits blackbody radiation. In reality, the location of the photosphere, $\tau_{\nu}\approx 1$, varies with the frequency because bound-bound transitions dominate the opacity
\citep[e.g.,][]{Tanaka2020}.
Nevertheless, this approximation may be valid at the early photospheric phase because the spectra are reasonably fitted by the Planck function with a single temperature.
Thus, we determine the velocity $v_{\rm phot}$ and temperature $T_{\gamma}$ of the photosphere by fitting the observed spectrum of AT2017gfo as 
\begin{equation}
    F(\lambda) = \frac{1}{D^2}\int^{r_{\rm max}}_{0}2\pi x B(\lambda, T(x))dx
\end{equation}
where $r_{\rm max}=v_{\rm max}t$ is the maximum radius at each epoch,  $B(\lambda, T)$ is the Planck function, and $D=40.7\,{\rm Mpc}$ is the distance from the earth \citep{2018Cantiello}. Here the relativistic Doppler effect is taken into account in an approximate manner.
Taking $\theta$ as $\cos\theta=\sqrt{1-(x/r)^2}$, $T(x)$ is
\begin{equation}
    T(x) = \frac{T}{\gamma [1-(v/c)\cos\theta]},
\end{equation}
where $\gamma$ is the Lorentz factor.

\begin{table}[]
    \centering
    \begin{tabular}{|c|c|c|}
        \hline
        Epoch [days] & $T_{\gamma}$ [K] & $v_{\rm phot}/c$ \\ 
         \hline
        1.43 & 4400 & 0.245 \\
        2.42 & 3150 & 0.22 \\
        3.41 & 2750 & 0.19 \\
        4.40 & 2600 & 0.155 \\
        \hline
    \end{tabular}
    \caption{Temperatures and velocities of photosphere obtained by fitting.}
    \label{tab:Tphot_vphot}
\end{table}

We obtain good fits for the spectrum at 1.43-4.40 days. Table~\ref{tab:Tphot_vphot} shows the temperatures and velocities of photospheres. In the following, we assume that the material above the photosphere is isothermal and the electron temperature $T_{e}$ equals $T_{\gamma}$. The assumption does not affect the result significantly as we are interested in the region close to the photosphere. 
\subsection{Rate equations and Atomic data}
For He,  21 levels (19 neutral levels up to $n=4$ for both singlets and triplets, one level for He\,II, and one level for He\,III) are considered for calculation.  
Natural transition rates (Einstein's A-coefficients) are taken from the NIST database \citep{NIST_Kramida}\footnote{\href{https://www.nist.gov/pml/atomic-spectra-database}{https://www.nist.gov/pml/atomic-spectra-database}}. The stimulated transition rates are calculated from the A-coefficients, photospheric temperature $T_{\gamma}$, and the geometric dilution factor $W(v)$ \citep{1978Mihalas}:
\begin{equation}
    W(v) = 0.5\Biggl(1-\sqrt{1-\biggl(\frac{v_{\rm phot}}{v}\biggr)^2}\Biggr).
\end{equation}
The line trapping of photons is given by the Sobolev escape probability $P(\tau_s)=[1-\exp(-\tau_s)]/\tau_s$ \citep{1970Castor}, where $\tau_s$ is the Sobolev optical depth (equation~\ref{eq:Sobolev}).
The photoionization cross sections from each level are obtained from \citet{2010Nahar}.
Excitations by electron collisions are considered \citep{1987Berrington, 2008Ralchenko}. 
Ionization by collisions of thermal electrons is considered but found to be negligible. Ionization by non-thermal particles from each level is considered as in equation~\ref{eq:ionization}. 
We assume that photoionization or recombination results in the ground level for singly and doubly ionized He. For He\,I, we treat direct recombination to the ground and excited levels separately. We inject 75\% of the recombination to excited states into the $2^3{\rm S}$ and the other 25\% into the $2^1{\rm S}$ levels. We also tried injection to $n=4$ levels and have confirmed that the result did not show any observable difference. We have expected this since natural transition cascades occur very quickly ($\sim 10^{6}\,\mathrm{s^{-1}}$) compared to other processes, such as photoionization. Therefore, all the triplet He atoms anyway decay to the $2^{3}{\rm S}$ level. 
Nonthermal excitation to the spin-triplet states is insignificant as singly ionized He dominates over neutral He, and recombination determines the population.

For Sr, we consider ten levels: five for Sr\,II ($\rm 5S_{1/2}, 4D_{3/2}, 4D_{5/2}, 5P_{1/2}, 5P_{3/2}$) and the other five for other ionization states (Sr\,I, Sr\,III, Sr\,IV, Sr\,V, Sr\,VI). The photon transition rates are taken from the NIST database \citep{NIST_Kramida}. 
We assume that all the collisional strengths are unity. This approximation does not cause serious errors in the level population because 
transitions are dominated by photons. Electron collisions start affecting the level population if the collisional strengths are higher than 100, which is unlikely. For more details, see Section~\ref{sec:atomic}.

\subsection{Spectrum calculation}
\label{sec:spectrum}

We follow the methodology presented by \citet{1990JefferyBranch} for line formation. Here, we outline the procedure.
We use a cylindrical coordinate in which the center of the merger ejecta is at the origin, and an observer is at $z=\infty$. The remaining two dimensions are described by a polar coordinate $(p, \theta)$, although we barely use $\theta$ as the ejecta is also axisymmetric. The z coordinate corresponds to the Doppler velocity since the ejecta coordinate satisfies $\vec{r}=\vec{v}t$ (free expansion of a point-source explosion).

We have assumed that the photosphere at $v_{\rm phot}$ is emitting blackbody radiation with a temperature of $T_{\rm phot}$. Outside the photosphere, the ejecta is optically thin except for the He or Sr lines. A photon with wavelength $\lambda$ can be scattered by the line transition with $\lambda_0$ at $z(\lambda) = v_z(\lambda)t$ that satisfies 
\begin{equation}
    \lambda_0=\lambda(1+v_z(\lambda)/c).
\end{equation}
Note that the treatment is to the first order of the Doppler effect. We discuss the impact of special relativity in Section~\ref{sec:relativity}.

Photons with wavelength $\lambda$ come either (i) directly from  the photosphere or (ii) after an interaction on the surface at $z(\lambda)=v_z(\lambda)t$. In case (ii), photons with wavelength $\lambda_0$ (in the ejecta-comoving frame) are scattered by the line transition, and they are observed as photons of $\lambda$  due to  the Doppler effect.  The sum of the two contributions gives the intensity from a point $(p, z)$:
\begin{equation}
    I(p,z)=\left\{
\begin{array}{ll}
S(p,z)(1-e^{-\tau})+B_{\nu}(T_{\rm phot})e^{-\tau} & (|p| < v_{\rm phot}t)\\
S(p,z)(1-e^{-\tau}) & ({\rm otherwise})
\end{array}
\right.
\end{equation}
where the Planck function  and the source function terms show the photospheric and the line scattering contributions, respectively. The source function $S(p,z)$ is given by
\begin{equation}
    S(p,z) = \frac{2h\nu^3}{c^2}\biggl(\frac{g_un_l}{g_ln_u}-1\biggr)^{-1}.
\end{equation}
 The total flux from $z$ is obtained by
\begin{equation}
    F(z) = \int_0^{p_{\rm max}} 2\pi p I(p,z) dp.
\end{equation}

We assume that a single line represents the line absorption for both He and Sr cases. The Sr triplet lines are separated by $\Delta \lambda = 882\,{\rm \AA}$, making $\Delta \lambda / \lambda \sim 9 \%$. Thus, this approximation does not affect the emergent spectral shape significantly, as the observed blueshifts ($\sim 0.2\,c$) and the width of the line feature are wider.

\section{Results} \label{sec:results}

\subsection{He model}
\label{sec:He}

\begin{figure*}[ht!]
    \centering
    \includegraphics[width=\columnwidth]{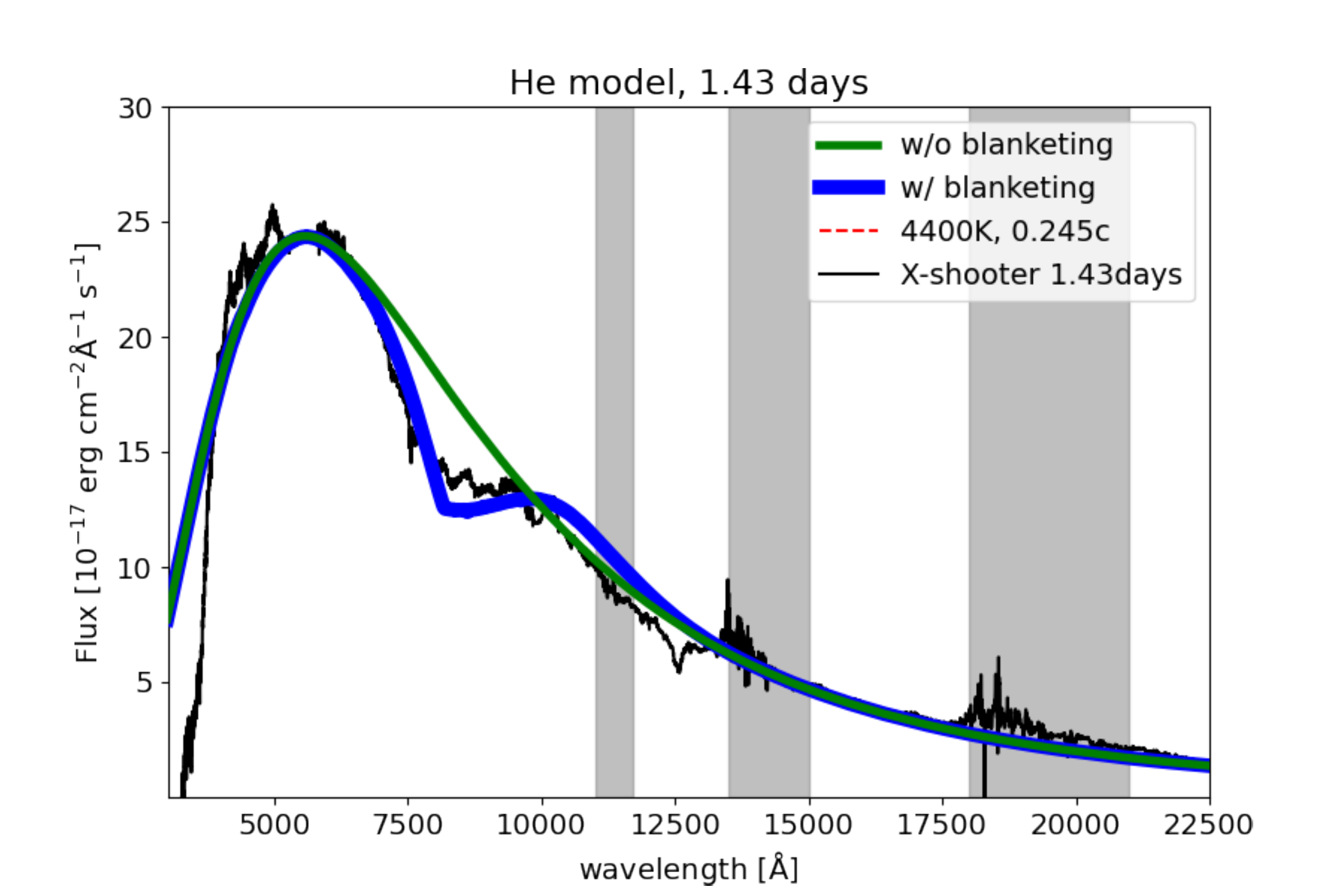}
    \includegraphics[width=\columnwidth]{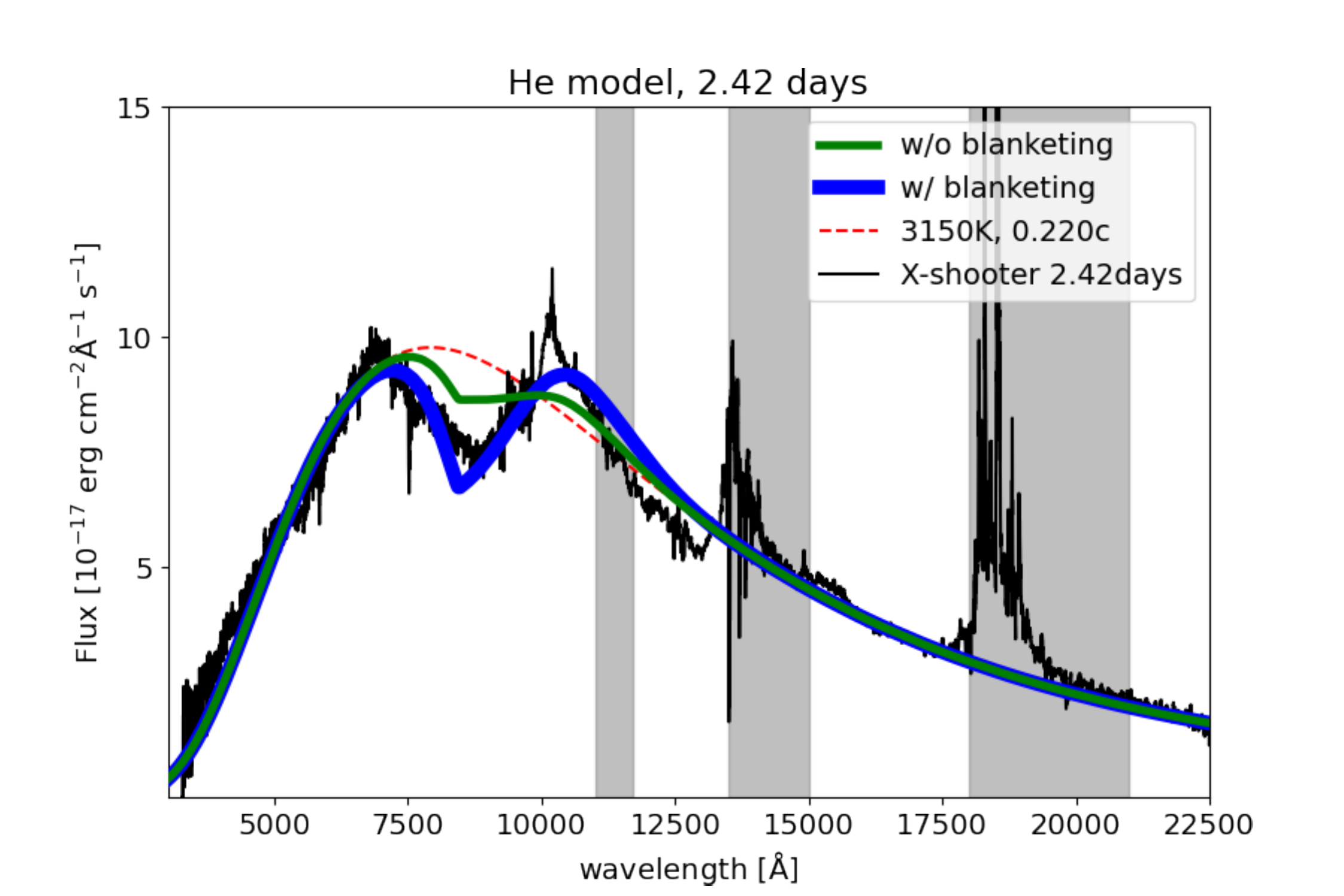}
    \includegraphics[width=\columnwidth]{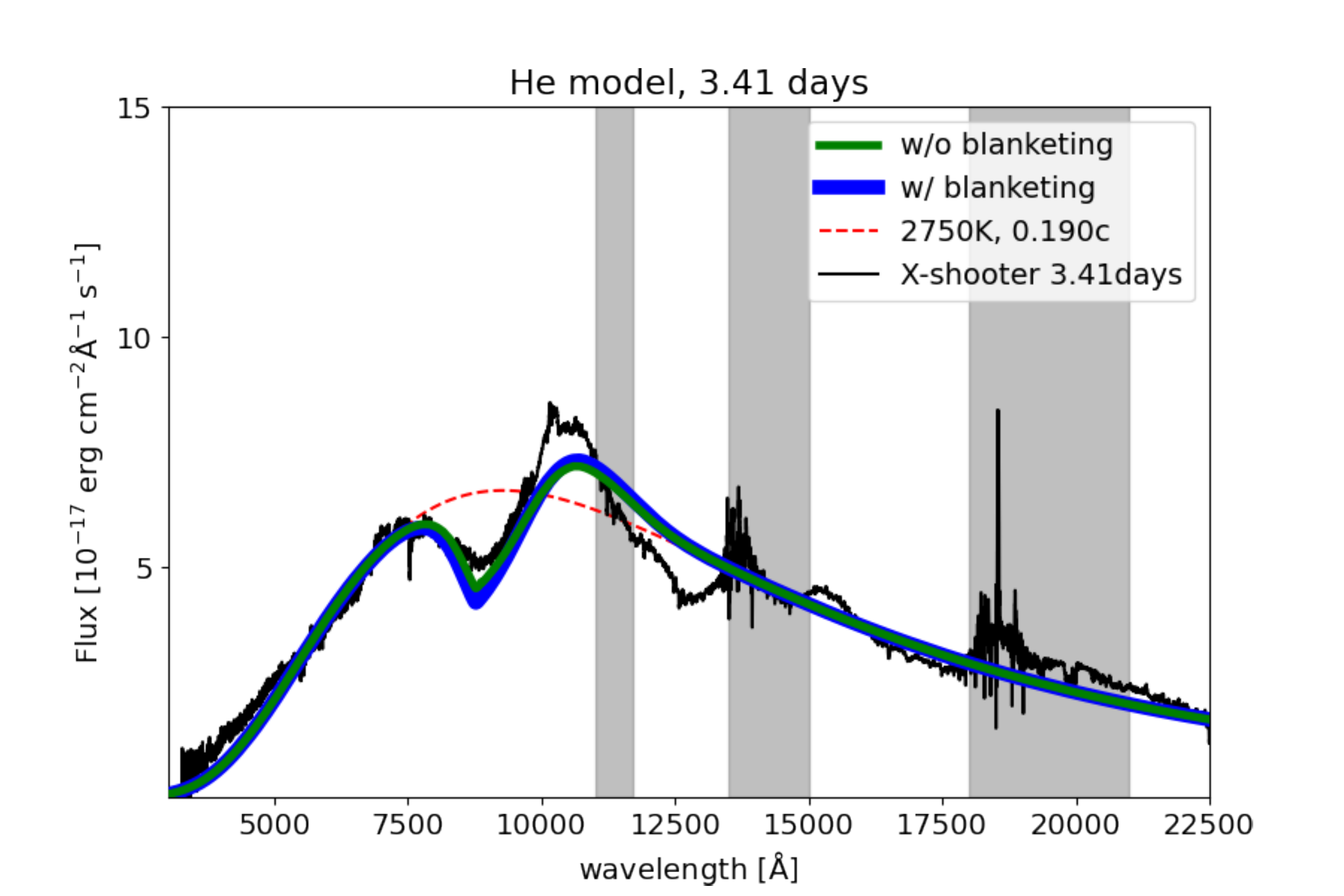}
    \includegraphics[width=\columnwidth]{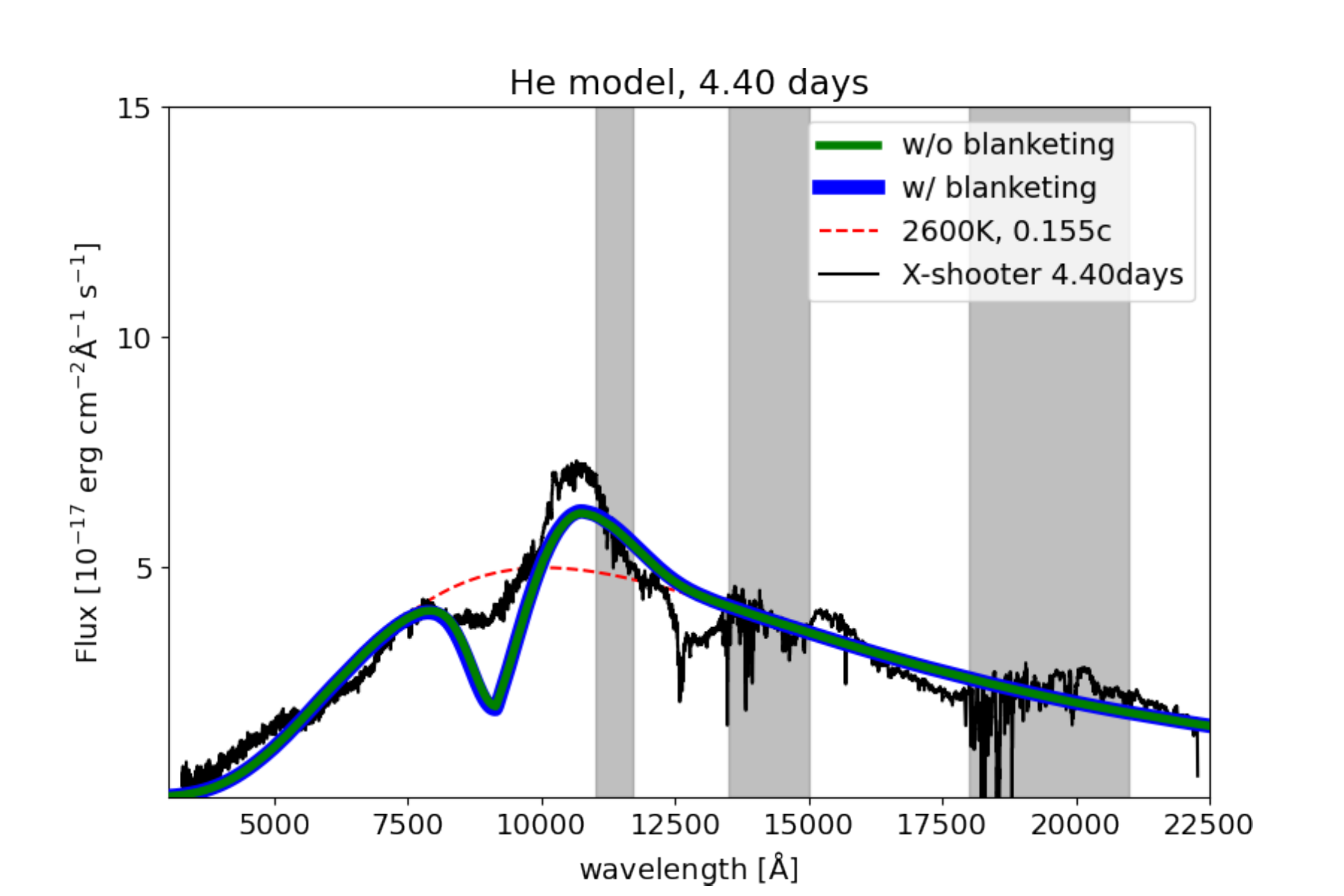}
    \caption{Comparison of He model spectrum to observed spectrum at 1.43-4.40 days from merger. The main model assumes suppression of photoionization due to metal line blanketing.
    The gray shaded regions show the wavelength where telluric absorption is significant. Spectroscopic data is obtained by \citet{Pian2017, Smartt2017}.}
    \label{fig:spectrum_He}
\end{figure*}

Figure~\ref{fig:spectrum_He} compares our model calculations to observed spectra. Our He model reproduces the absorption feature at 3.41-4.40 days, while the prediction for the first two epochs is weaker than the observations. The model spectra match well if photoionization from the $n=2$ states are ignored for the first two epochs. Photoionization has a minor effect for 3.41 and 4.40 days as electron collision replaces photoionization. 
Therefore, He of $8\times 10^{-5} \Msun$, which corresponds to 0.2\% of the total ejecta, can naturally produce the line feature observed in AT2017gfo. 

The spectral feature persists for the first 3.4 days and gets stronger at 4.4 days. The estimates of equation~\ref{eq:23S} suggest that the line optical depth gets weaker as time at each velocity coordinate. However, the photospheric velocity decreases with time. Therefore, the decrease in photospheric density is slighter than $t^{-3}$ in our calculation.

Photoionization significantly contributes to the depopulation of the $2^{3}{\rm S}$ level for the first $2.5$ days. We have found that the photoionization rates should be 0--0.3\% and 0--30\% of blackbody values for 1.43 and 2.42 days to reproduce the observations (see Section~\ref{sec:UV and 23S}). The ionization threshold is $4.8\,{\rm eV}$, corresponding to $\sim 2600$ \AA\ in the ejecta rest frame. In the  model with photoionization, we assume that the blackbody emission continues to this wavelength. However, the spectrum at 1.43 days shows a sharp cut-off feature at a wavelength shorter than $4000$\,\AA  (figure~\ref{fig:spectrum_He}). Therefore, we expect that our model without the blanketing significantly overestimates the photoionization rate at 1.43 days. A strong blanketing is naturally expected because heavy ions in the ejecta absorb UV photons more efficiently than optical photons. The suppression of near-UV emission by a factor of a few hundred is typically seen in the results of radiative transfer simulations (\citealt{Domoto2021}).
To properly assess UV blanketing, knowledge of UV absorptive opacity \citep{1978Mihalas} of {\it r}-process elements is required.  

\begin{figure*}[ht!]
    \centering
    \includegraphics[width=\columnwidth]{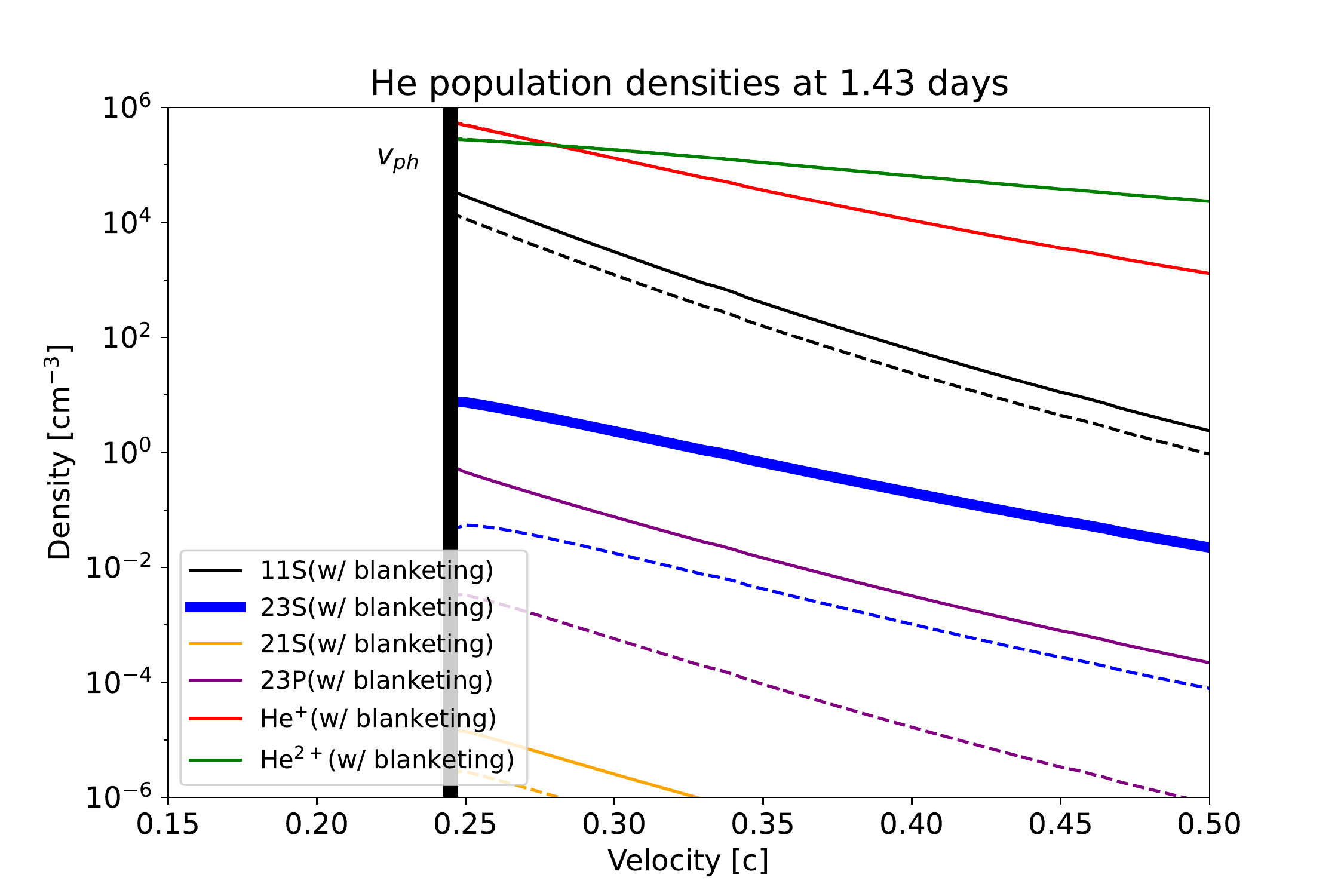}
    \includegraphics[width=\columnwidth]{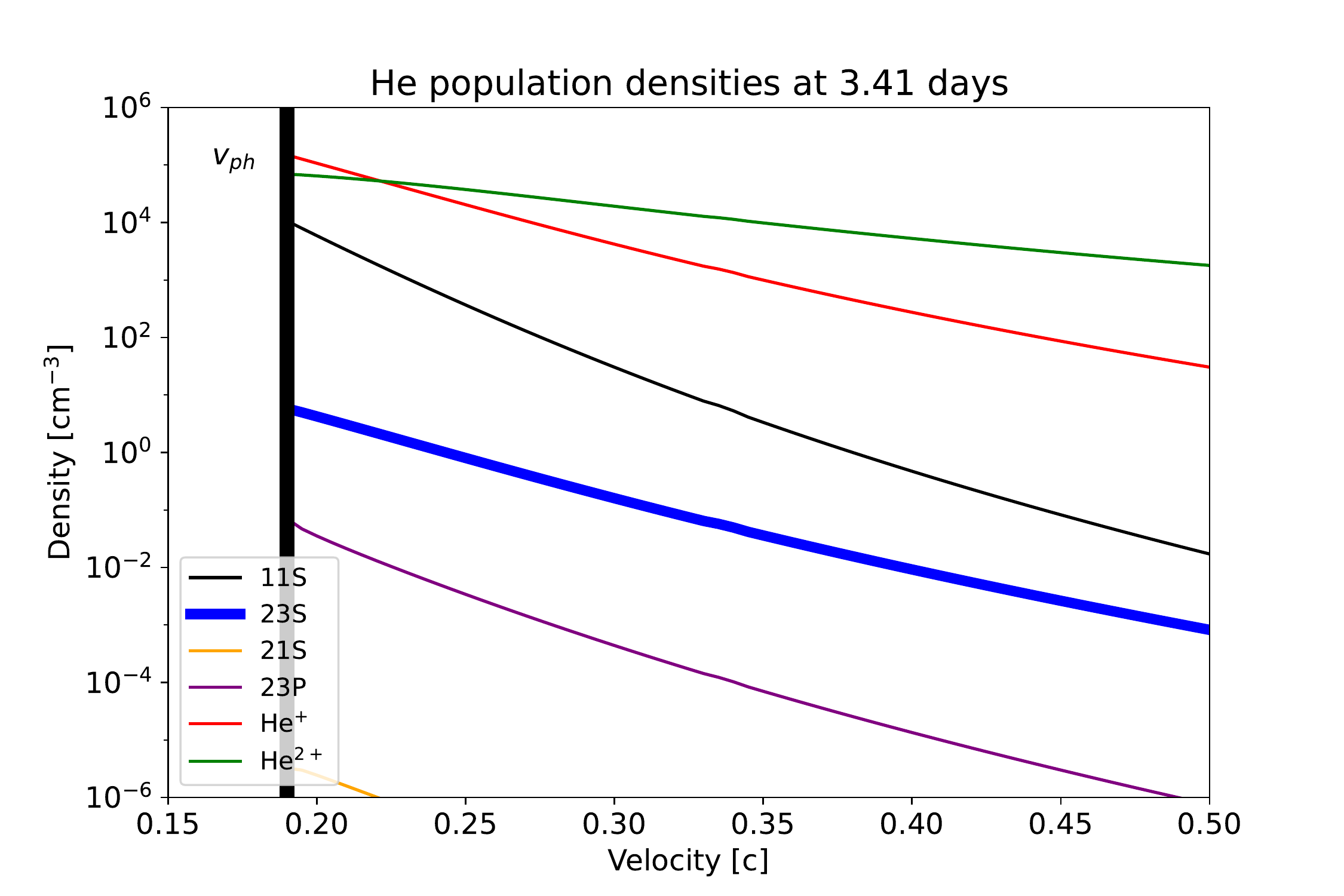}
    \caption{Population of excited He levels against velocity shell at 1.43 days on the left and 3.41 days on the right. 
    Only outside the photosphere is shown, where the photospheric velocities are $0.245c$ and $0.19c$ at $1.43\,{\rm days}$ and $3.41\,{\rm days}$, respectively.  The spectral line feature appears when $n({\rm He\,I,2^3{\rm S}})\gtrsim 7.4t_{d}^{-1}\,{\rm cm^{-3}}$ around the photosphere is satisfied. The solid lines show the population in the case that the photoionization is ignored. The dotted lines for the 1.43 days model show population without the blanketing. Note that
    the line blanketing does not affect the level population at $3.41$ days.}
    \label{fig:He Population}
\end{figure*}

Figure~\ref{fig:He Population} shows the population profile of each level at 1.43 and 3.41 days. The population of $2^{3}{\rm S}$  at the photosphere is higher than the critical value (see equation \ref{eq:HeI_crit}). For the first epoch, line blanketing for UV photons needs to be considered: no blanketing model (the dashed line) predicts two dexes lower abundance, which is insufficient for line formation. The significant difference between ${\rm He}^{+}$ and  $2^{3}{\rm S}$  is due to efficient photoionization. The next well-populated level is $2^{3}{\rm P}$ which is excited from $2^{3}{\rm S}$ by blackbody photons. The ratio between $n_{2^{3}{\rm S}}$ and $n_{2^{3}{\rm P}}$ is well described by
\begin{equation}
    \frac{n_{2^{3}{\rm P}}}{n_{2^{3}{\rm S}}} = \frac{3}{1}\exp\biggl[\frac{-(E_{2^{3}{\rm P}}-E_{2^{3}{\rm S}})}{kT_{\rm phot}}\biggr]\times W(v),
\end{equation}
where $T_{\rm phot}$ is the photospheric temperature.
The ratio indicates that the blackbody photons govern the transition between $2^{3}{\rm S}$ and the $2^{3}{\rm P}$. Even the second most populated excited level, $2^{3}{\rm P}$, has somewhat lower densities  $\lesssim 1\,{\rm cm^{-3}}$.
The absorption lines at $5877{\rm \AA}$ and $7067\,{\rm \AA}$ are absent because of the small population.
All the other excited levels have densities less than $10^{-2}\,{\rm cm^{-3}}$, which is too small to produce an observable effect on the spectrum. Therefore, we expect that no He line other than $10833{\rm \AA}$ appears in kN spectra.

\subsection{Sr model}
\label{sec:Sr}

\begin{figure*}[htbp]
    \centering
    \includegraphics[width=\columnwidth]{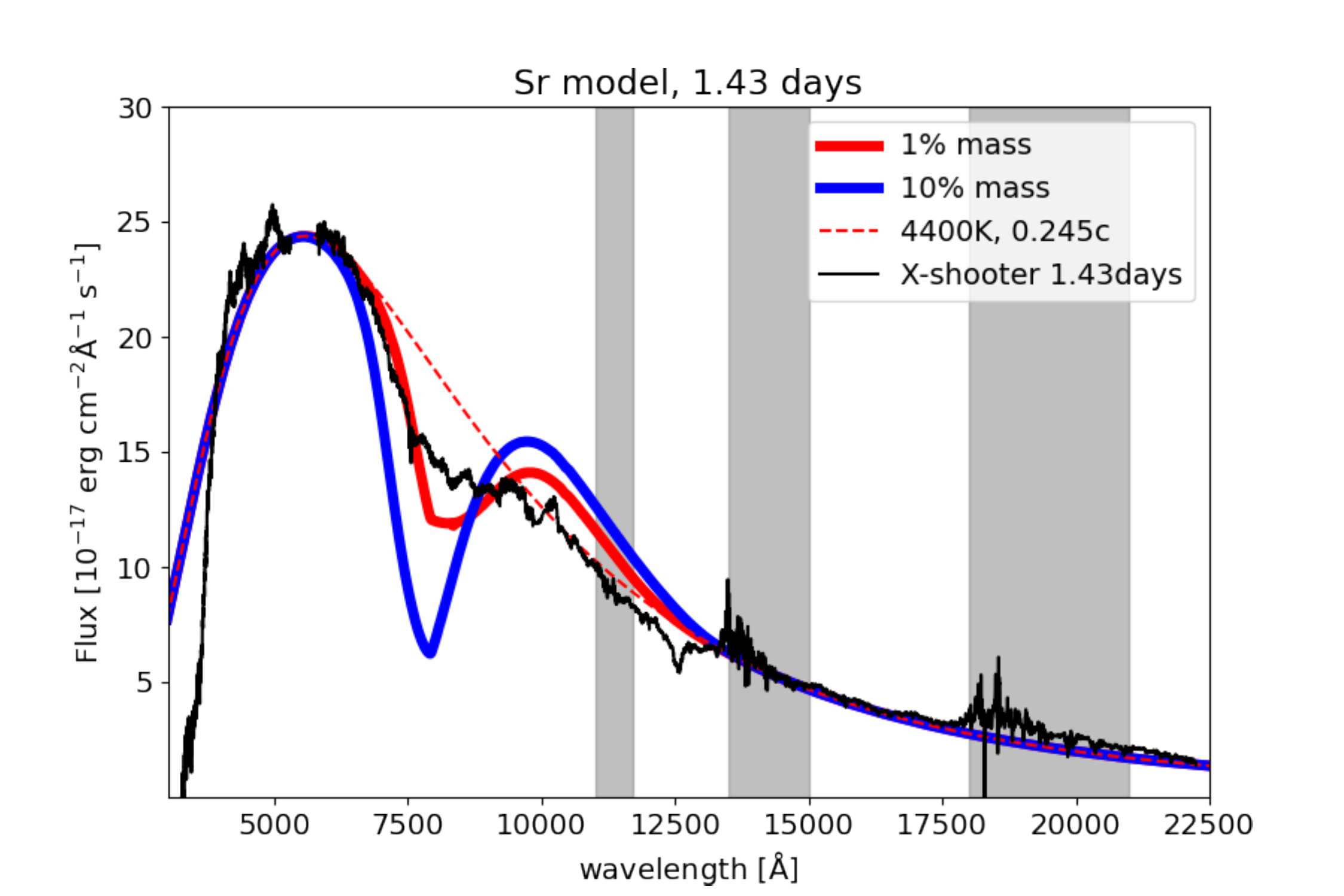}
    \includegraphics[width=\columnwidth]{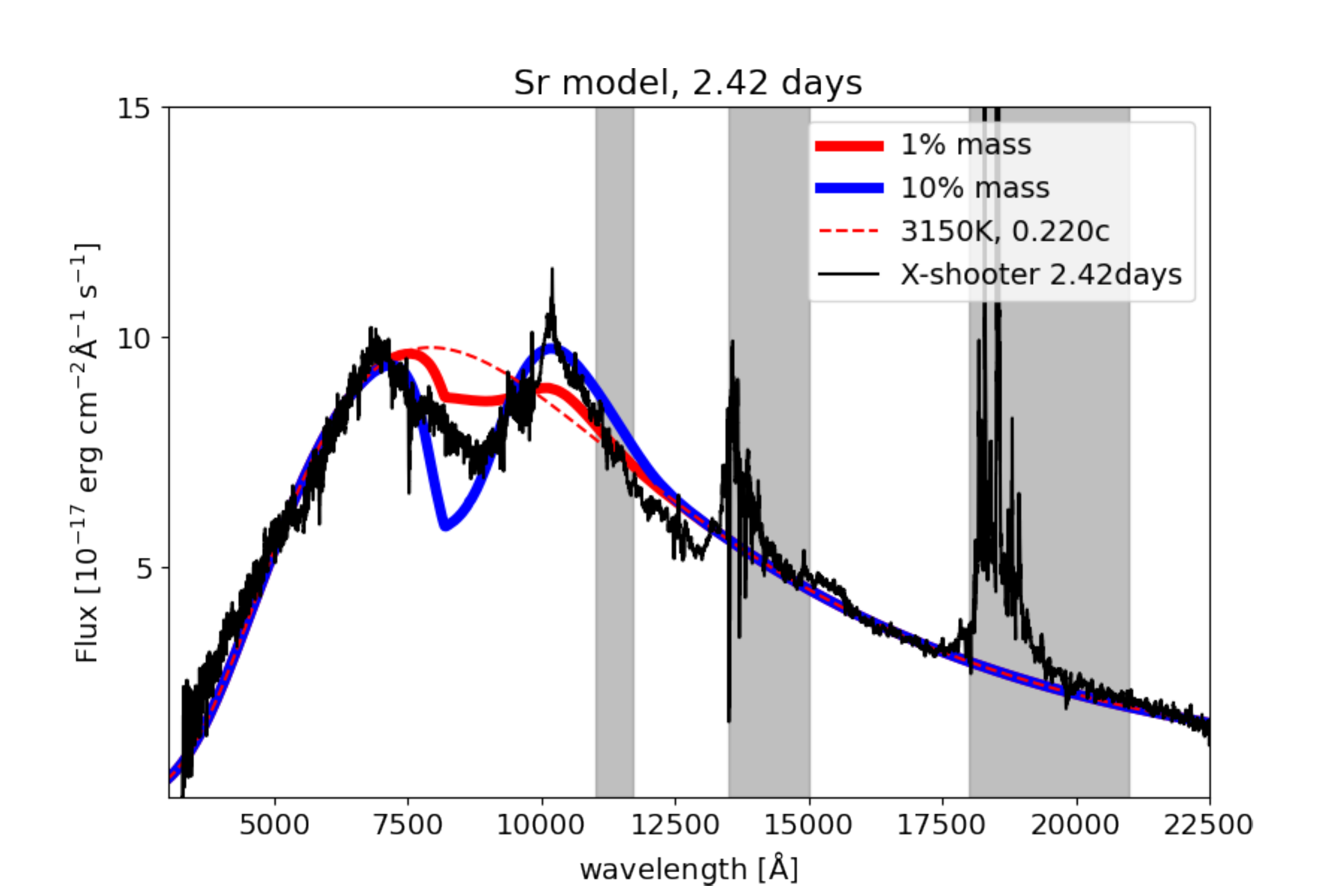}
    \includegraphics[width=\columnwidth]{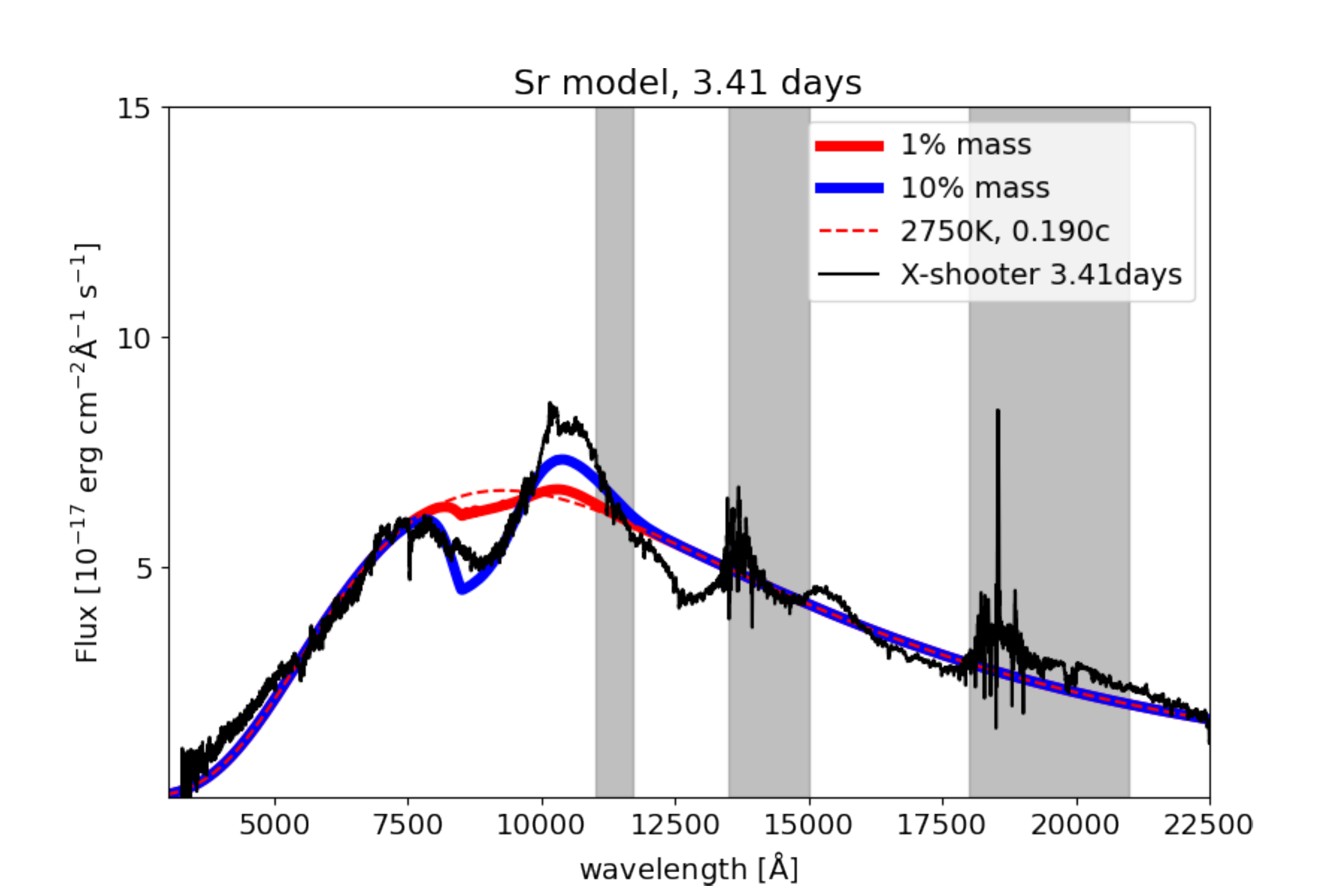}
    \includegraphics[width=\columnwidth]{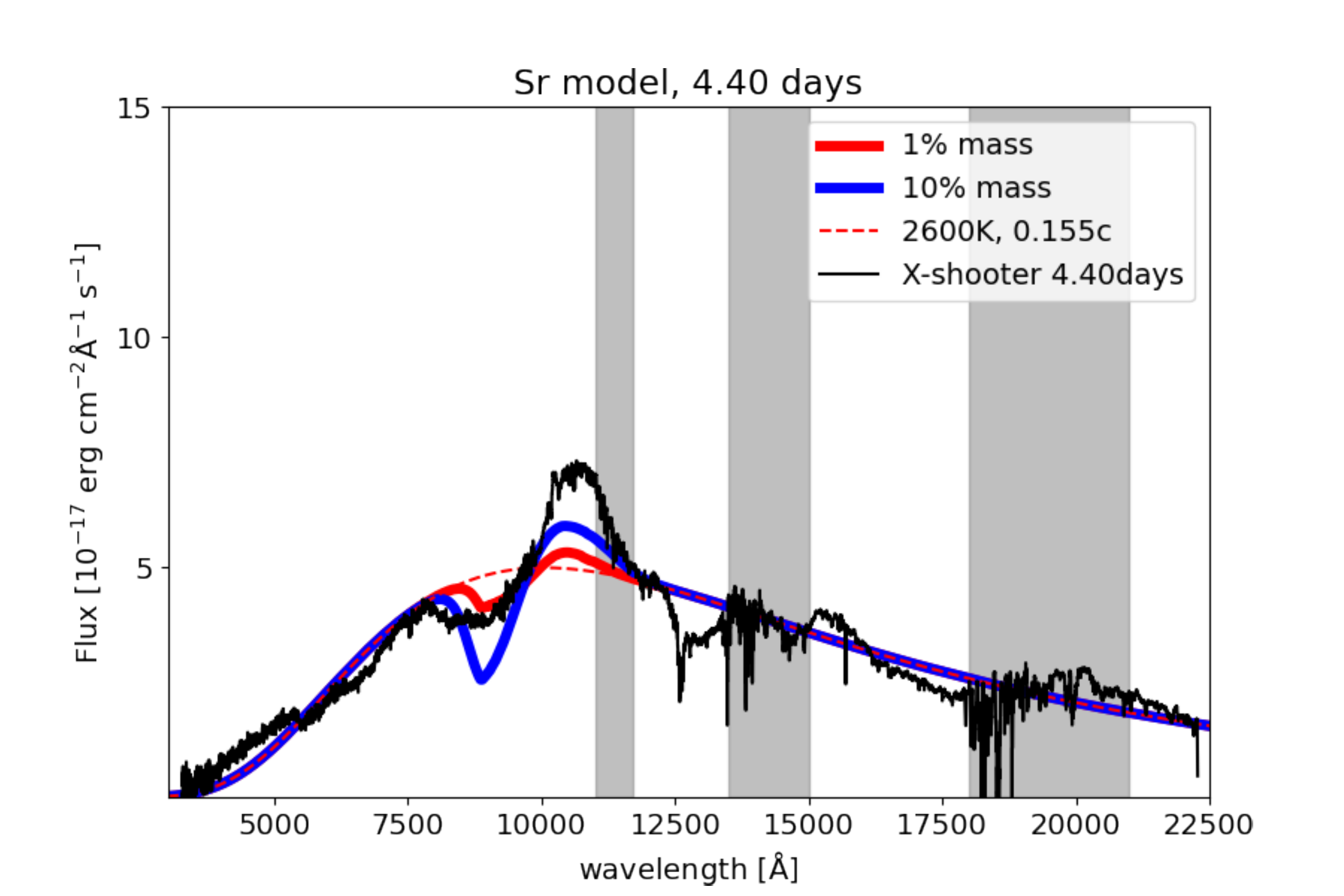}
    \caption{Same as figure \ref{fig:spectrum_He} but for the Sr model.  The red and blue lines represent model predictions with Sr mass fractions of 1\% and 10\%.}
    \label{fig:spectrum_Sr}
\end{figure*}

In figure~\ref{fig:spectrum_Sr}, we show the results for the Sr model. In the fiducial model (mass fraction of 1\%, red lines), the P-Cygni feature is stronger than the observation at 1.43 days, while it gets weaker at later epochs. The enhanced model (mass fraction of 10\%, blue lines) produces features consistent with observations at 2.42-4.40 days. However, in the first epoch, the feature is too strong.

The decline of the line strength results from overionization by nonthermal particles. Note that the decline is quicker than He as in the estimates in equations~\ref{eq:23S} and \ref{eq:Sr population}. If Sr\,II is the line's origin, the photosphere must go inside quickly from 1.43 to 4.40 days. 
Another intriguing possibility is that the spectral feature at the first two epochs is from Sr\,II, and He\,I takes over in the later epochs. Constraints on the time evolution of the photospheric density are crucial for further modeling.

Contrary to He, the population ratios between different energy levels of singly ionized Sr are not much affected by the nonthermal particles because the 4D levels are in pseudo-equilibrium with the ground level via photon (allowed) transitions to the 5P levels. 
The photospheric temperature determines the population ratio of 4D to the ground level. The decrease in temperature lowers the ratio ($n_{\rm Sr^{+}_{4D}}/n_{\rm Sr^{+}_{5S}}$). 

At temperatures around $3000-4000$ K, lower temperature favors Sr to be singly ionized in Saha equilibrium. This effect  maintains a somewhat large population of  4D  in the later epochs in  LTE analyses. However, in the non-LTE case, the dependence of ionization structure on local temperature becomes significantly weak. Nonthermal ionization is not affected by the temperature, in contrast to the ionization by thermal photons, which strongly depends on the temperature via the exponential factor. Although the recombination coefficient introduces a weak dependence, the overall ionization structure does not depend much on temperature. The 4D population now depends strongly on the local density, which affects recombination rates.

\begin{figure*}[ht!]
    \centering
    \includegraphics[width=\columnwidth]{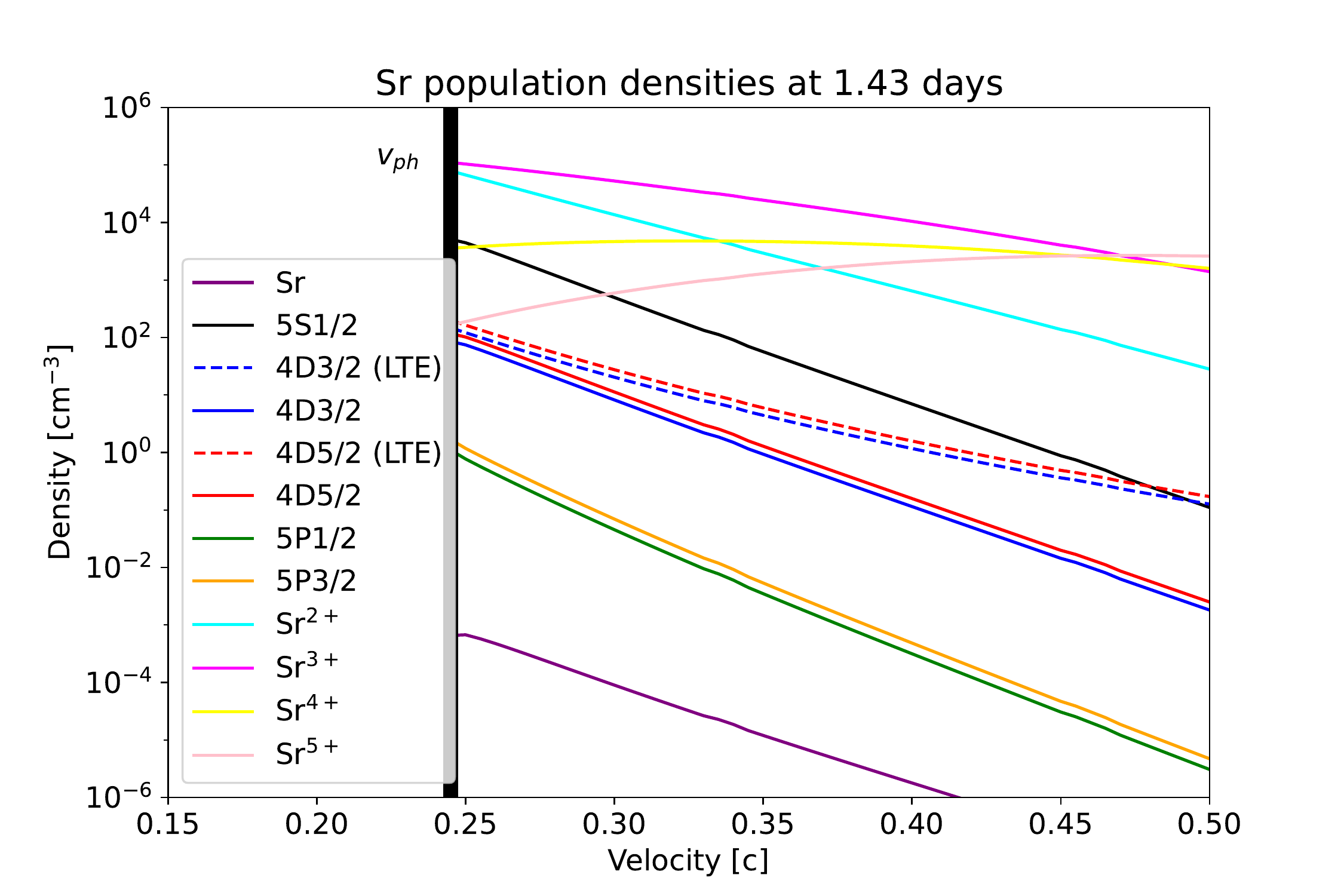}
    \includegraphics[width=\columnwidth]{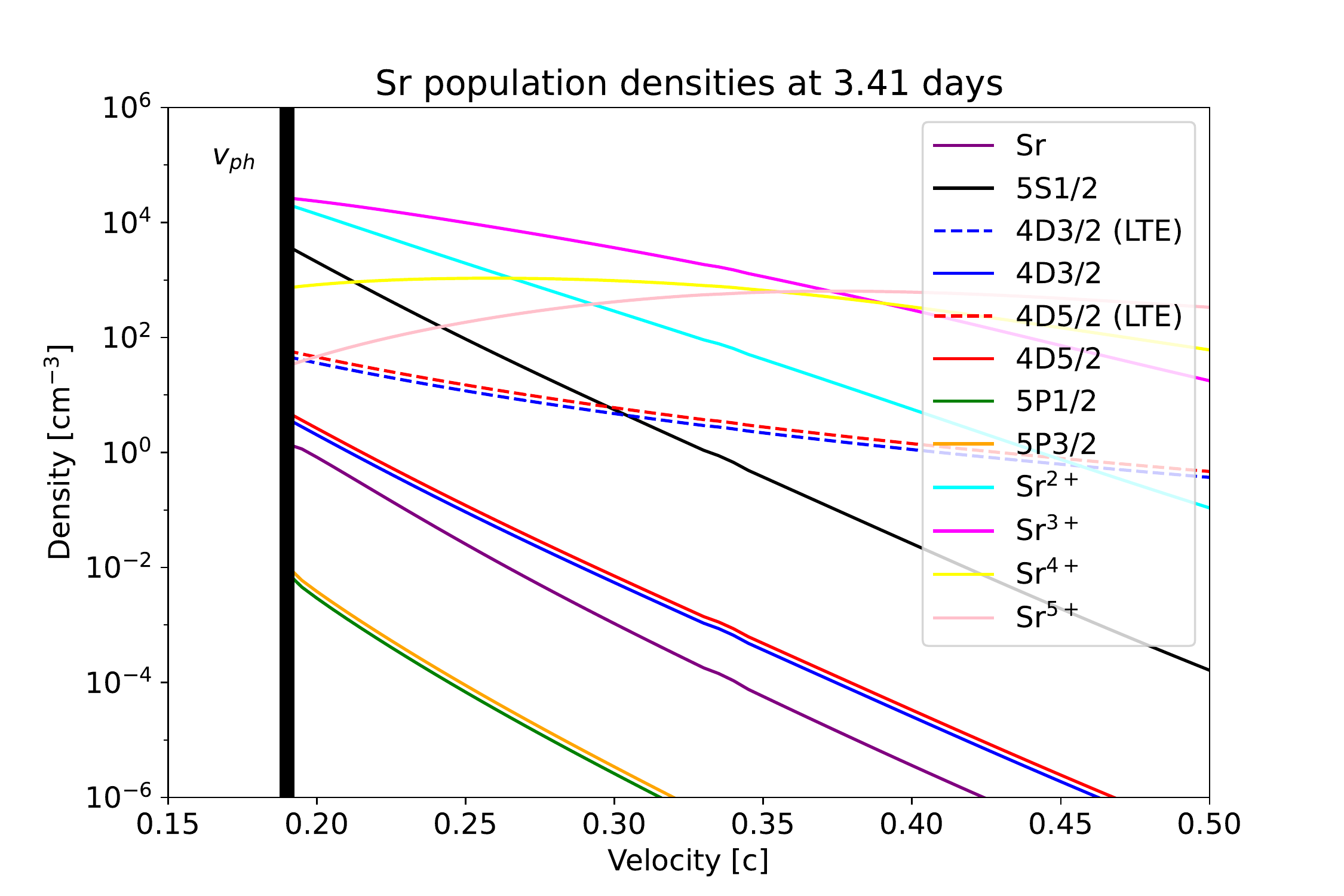}
    \caption{Same as \ref{fig:He Population} but for the Sr model. The Sr II line feature appears when $n({\rm Sr\,II,4^2{\rm D}})\gtrsim 50t_{d}^{-1}\,{\rm cm^{-3}}$ is satisfied. 
    Dashed lines represent populations for $4^{2}$D  assuming LTE of the photospheric temperature.
    }
    \label{fig:Sr Population}
\end{figure*}

\subsection{Comparison to SN Ib on He lines}
\label{sec:compare to Ib}
Interestingly, in the kN condition, the level populations for the singlet levels are far less populated than the triplet states. The fact that the singlet populations are lower than that of the triplet states by a huge factor is in stark contrast with the case of type-Ib supernovae (SNe), which show spectral features also from the singlet  levels such as the $20589{\rm \AA}$ line \citep{1991Lucy}. 

\begin{figure}
    \centering
    \includegraphics[width=\columnwidth]{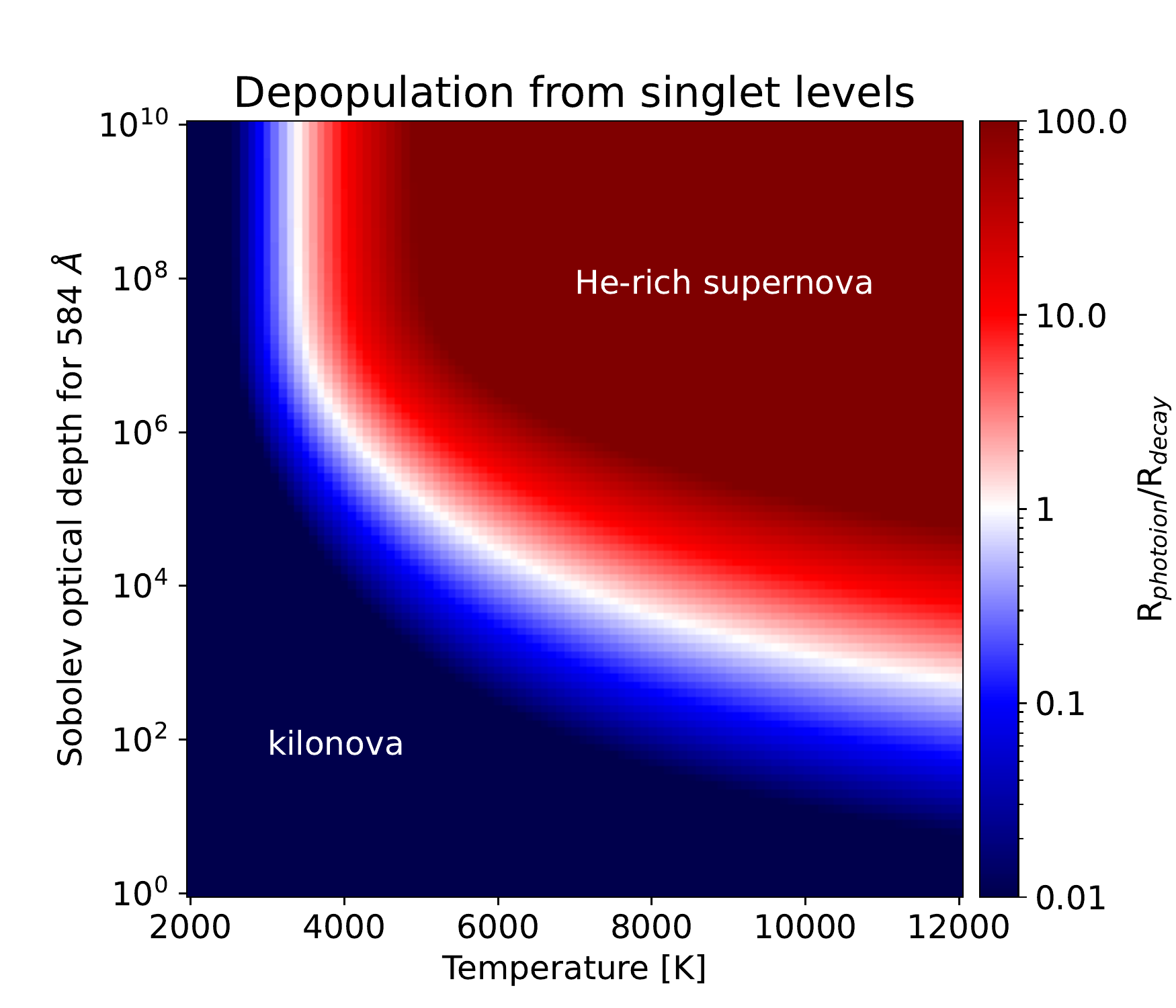}
    \caption{Ratio of photoionization rate and the decay rate to the ground state from singlet excited states $2^{1}{\rm S}$ and $2^{1}{\rm P}$. High and low ratios mean that the depopulation dominantly occurs via photoionization and natural decay, respectively.}
    \label{fig:singlet_depopulation}
\end{figure}

Both the singlet and triplet states are populated via recombination from singly ionized He. However, the depopulation mechanism could make the difference. Figure~\ref{fig:singlet_depopulation} shows the ratio of the photoionization rate and natural decay rate on the temperature-optical depth (i.e., neutral He density) plane. Here, we assume that the population of $2^{1}{\rm P}$ is
\begin{equation}
\frac{n_{2^{1}{\rm P}}}{n_{2^{1}{\rm S}}} = \frac{3}{1}\exp\biggl[\frac{-(E_{2^{1}{\rm P}}-E_{2^{1}{\rm S}})}{kT_{\rm phot}}\biggr]\times W,   
\end{equation}
and $W = 0.5$ (dilution factor at the photosphere).
Photoionization is the dominant depopulating mechanism for type-Ib supernovae for the singlet levels $2^1{\rm S}$ and $2^1{\rm P}$. The crucial factors are the high density of neutral He and the high photospheric temperature. The Sobolev escape probability is so low that a photon emitted by the allowed transition between $2^{1}{\rm P}$ and $1^{1}{\rm S}$ immediately excites He from the ground level. In this case, triplet and singlet levels are populated comparably. On the other hand, the He density in kNe ejecta is not sufficient to hinder the natural decay from the ${2^{1}{\rm P}}$ to the ground levels. Therefore, once a $2^{1}{\rm S}$ atom is excited to ${2^{1}{\rm P}}$, it quickly decays to the ground level via the very efficient ($A_{\rm 21P\rightarrow 11S} = 2.0\times 10^{9}\,{\rm s}^{-1}$) natural transition.

\section{Discussion}

\begin{table*}[]
    \centering
    \begin{tabular}{lll}
        \hline \hline
        Literature & Type & $X_{\rm He}$ \\
         \hline
        \cite{Goriely2015MNRAS} & Dynamical ejecta & $\sim 10^{-1}$\\
        \cite{Lippuner2017MNRAS} & Massive neutron star wind & $\sim 10^{-1}$ \\
        \cite{Kawaguchi2022ApJ} & Massive neutron star MHD wind & $10^{-2}$--$10^{-1}$ \\
        \cite{Fernandez2013} & Black-hole disk wind & 
        $\sim 10^{-2}$ \\
        \cite{Kullmann2022MNRAS} & Dynamical ejecta & $\sim 10^{-2}$ \\
        \hline
        \cite{2022Perego} & Dynamical ejecta & $\sim 10^{-3}$ \\
        \cite{2018Wanajo} & Phenomenological model & 
        $10^{-4}$--$10^{-3}$\\
        \cite{2022Fujibayashi} & Dynamical~\&~post-merger ejecta & 
        $10^{-4}$--$10^{-3}$ \\
        \cite{2022Perego} & Spiral-shock wind & $\sim 10^{-4}$ \\
        \cite{Wanajo2014ApJ} & Dynamical ejecta & $\sim 10^{-5}$ \\
        \hline \hline
    \end{tabular}
    \caption{Simulated mass fraction of Helium in neutron star merger ejecta in the literature. Note that the values depend on mass ratio and other conditions. Therefore, values should be used only as a rough guide to the literature. For the exact mass fractions, readers should refer to original papers.}
    \label{tab:Heproduction}
\end{table*}

\subsection{He production and spectral feature in merger ejecta}

We have shown that 0.2\% of He is sufficient for forming the P-Cygni profile. Non-LTE effect significantly affects the population at the line-forming region and the line strengths even in the first 1.5 days.

He is copiously produced in the high-entropy component of neutron star merger ejecta due to the low density when it crosses the temperature range of He burning. \citet{Fernandez2013} derive a formula following the formulation by \citet{1997Hoffman}. Briefly, the bottleneck reactions for the He burning needed for the onset of the $\alpha$-process and the subsequent {\it r}-process is $^{4}{\rm He}(\alpha n, \gamma)^{9}{\rm Be}(\alpha, n)^{12}{\rm C}$. Therefore, the fraction of He at the end of the $\alpha$-process can be estimated by considering the integrated reaction rate from $T=5\cdot 10^9\,{\rm K}$ to $2.5\cdot 10^9\,{\rm K}$ when the burning of He nuclei occurs. 
We can estimate the typical entropy of the ejecta from the He fraction. Assuming $X_{\rm He} = 2\times 10^{-3}$, 
\begin{equation}
    S \simeq 18\times \biggl(\frac{\bar{Z}}{36}\biggr)^{1/3}(1-2Y_e)^{1/3}\biggl(\frac{\tau_{\rm dyn}}{10 [{\rm ms}]}\biggr)^{1/3}\,[k_B\,{\rm nuc^{-1}}],
\end{equation}
where $\bar{Z}$, $Y_e$, $\tau_{\rm dyn}$ are the average proton number of the seed nuclei, electron fraction, and the dynamic timescale, which is the timescale for the temperature decrease by a factor of $e$ \citep{1997Hoffman}. The dynamic timescale $\tau_{\rm dyn}$ is uncertain, although \citet{Fernandez2013} estimate $\tau_{\rm dyn} \simeq 100\,[{\rm ms}]$ for the black-hole disk wind.

Table~\ref{tab:Heproduction} 
shows the estimated values of $X_{\rm He}$ in various nucleosynthesis results of merger ejecta in the literature\footnote{The nucleosynthesis results of merger simulations are usually shown for the entire ejecta in the literature. Here, however the abundances of the shells with velocity of $\gtrsim 0.2c$ are relevant, and thus, we should note that the values in Table \ref{tab:Heproduction} may not be  representative of the abundance in the line-forming region.}. It is worth noting that there are many simulations in which a somewhat large amount of He, $X_{\rm He}\sim 10^{-2}$, is produced. Thus, the explanation of the $8000{\rm \AA}$ feature by the He\,I line seems plausible.
If correct, because the predicted values of $X_{\rm He}$ depend on the ejecta entropy, spectroscopic observations of future kNe provide a better understanding of the mass ejection mechanism. 

\citet{2022Perego} discuss the He production in dynamical and spiral-wind ejecta and argue that He cannot produce line signature except in a suppressed luminosity model. They assume He masses of $1$ -- $10\cdot 10^{-6}\,\Msun$, an order of magnitude lower than our assumption ($8\times 10^{-5}\,\Msun$). They use an analytical approximation model developed for He-rich ejecta of double-detonation type-Ia SNe in \textsc{tardis}. However, as we have shown in Section~\ref{sec:compare to Ib}, non-LTE effects depend on the density and temperature 
of the ejecta. Therefore, it is unclear whether the model applies to the case of a kN.

\subsection{Non-LTE effect on Sr}

The non-thermal particles hinder the recombination of Sr\,III from 1.43 days to later epochs. \citet{2019Watson} model the line signature assuming the Sr\,II lines are the origin. Using a 1D Monte-Carlo radiative transfer code \textsc{tardis}, they argue that $5 \times 10^{-5}\,\Msun$ of Sr is needed in the optically thin part for $1.43$ days, whereas $\sim 1 \times 10^{-5}\,\Msun$ is sufficient for $2.42$ -- $4.40$ days. The decrease in the required amount of Sr is likely a consequence of the recombination from Sr\,III to Sr\,II. In our non-LTE model, non-thermal electrons hinder recombination. As a result, the line signature gets weaker over the first few days. \citet{Gillanders2022} model the signature using the same code. The mass of Sr\,II in the line-forming region is estimated to be $0.1$ -- $10$ $\times 10^{-6}$ \Msun, which is consistent with our values. However, the estimated total ejecta mass in the line-forming region $v \gtrsim 0.12\,c$ is $\sim 10^{-4}\,\Msun$, which  is considerably low compared to the mass required to produce the sufficient radioactive luminosity. This is likely a consequence of the LTE approximation. In our non-LTE analysis, on the other hand, we find the ejecta mass at $v \gtrsim 0.12\,c$ is $\sim 0.04M_{\odot}$ if the mass fraction of Sr is $1\%$.
\citet[Figure~2]{Domoto2022} show an LTE model with the line optical depth at 3.41 days remains as large as 1.43 days. We expect that the optical depth gets smaller at the later epoch due to the ionization by non-thermal particles.
Therefore, including the effect of non-thermal ionization is crucial to estimate the elemental mass in the line-forming region.

\subsection{Comparison of He and Sr line features}
\label{sec:early}

Both He and Sr could show the P-Cygni profile. It is crucial to distinguish between these two elements. Other transition lines of He and Sr would be clear evidence for the distinction. For He, however, no other lines could be used as a test because the expected populations are too low.

Singly ionized Sr has strong transitions at 4079 and 4217\,\AA\ between the 5S (ground) and 5P levels. The Sobolev optical depths for these transitions are always much higher than the Sr\,II triplet transitions because (i) the population at the ground level is higher than the 4D
levels by the Boltzmann factor ($\sim 10$ including degrees of freedom and assuming the temperature of $\sim 5000\,$K), and (ii) the oscillator strengths are higher by a factor of 8. The sharp UV decline in the spectrum at 1.43 days may be a consequence of this transition \citep{2019Watson}. These transitions are not present in spectra at 2.42-4.40 days. This seems to be against the Sr model, but we cannot firmly exclude it because the continuum emission at that wavelength is weak due to the low photospheric temperatures. 
Spectra at early (0.5-1.5 days) epochs would be crucial to infer these transitions.
Although there are a few spectroscopic observations at early epochs \citep{2017McCully, 2017Andreoni}, we could not model the strengths of the 5S-5P transitions because the shorter wavelength edge of the feature was not covered, and there are many line transitions at $\sim 4000$\,\AA. Near-UV ($\sim 2000$\,\AA) to near-IR spectra ($\sim 10000$\,\AA) at these early epochs will be useful to model the Sr\,II line feature as well as the He photoionization we discussed in Section~\ref{sec:UVflux}.

The two 4D levels of Sr can decay to the ground (5S) level by forbidden transitions. However, their emission lines are too weak to be observed. Blackbody photospheric emission dominates in early times. At later ($\gtrsim 10$ days) epochs, Sr is overionized by beta particles to higher ionization states. 
Assuming that the Sr mass fraction is 1\%, the level population is in thermal equilibrium of $3000~{\rm K}$, and 1\% of Sr is singly ionized at 10\,days, the luminosity of the line photon is $5\times 10^{34} {\rm erg\,s^{-1}}$. The flux is too low to be observed.

\subsection{Model uncertainties and limitations}

Our model assumes that the continuum emission is thermal radiation of the photosphere with a single velocity and temperature, ignoring the  emission, scattering, and absorption in the optically thin region apart from the bound-bound transition of He\,I and Sr\,II. The photospheric temperature and density, which are obtained from the observed spectrum, are used to determine the level population and ionization states of both elements. In reality, temperature may be different from the color temperature of the observed spectrum or the density structure is different from what we have assumed.
Here, we discuss the caveats in our modeling.

\subsubsection{Lines of other ions}
The kN ejecta are composed of heavy elements and their bound-bound transitions of these elements dominate the opacity. Our model implicitly assumes that the photosphere is determined by the line expansion opacity. However, the opacity varies with wavelength.
In the wavelength region of $\gtrsim 8000{\rm \AA}$, lanthanides and actinides likely to dominate the opacity, which might affect the interpretation \citep{Kasen2013ApJ,Tanaka2013ApJ,Tanaka2020,Fontes2020MNRAS,Fontes2022MNRAS}. Furthermore, several strong lines are expected to exist above the photosphere. In fact, \cite{Domoto2022} find that the strong lines of La III around $1.4{\rm \mu m}$ and Ce III around $1.7{\rm \mu m}$ produce the spectral features. 

\subsubsection{Asphericity and radial gradient in the abundances}
Our model assumes a spherically symmetric ejecta and uniform elemental abundances in the line-forming region for simplicity. In reality, however, the ejecta shape and the composition distribution are expected to be somewhat aspherical, which affects the light curve and spectrum (\citealt{Kawaguchi2018ApJ,Wollaeger2018MNRAS,Bulla2019MNRAS,Darbha2020ApJ,Korobkin2021ApJ}; see also \citealt{2023Spherical}, who argue that the ejecta is highly spherical from analyzing blackbody-like emission and the line profile).
Because our model focuses on the depth of the observed absorption feature, it is sensitive to the He and Sr abundances of the material moving toward an observer. 
Therefore, our model may over- or under-estimate the amount of He and Sr depending on the degree of the ejecta asphericity. 

The elemental abundances likely  vary with the velocity of the shells. Our finding that the Sr (He) line strength decreases (increases) with time suggests that the  a model that the Sr (He) abundance which decreases (increases) with the expansion velocity is more successful in reproducing the time evolution of the observed line shape.  Therefore, we will explore the effects of the asphericity and radial gradient  of the ejecta compositions in future works.

\subsubsection{UV flux and the photospheric temperature}
\label{sec:UVflux}

The flux of ionizing photons ($> 4.8\,{\rm eV}$) is the crucial parameter for He. Higher photospheric temperature results in higher UV flux. However, as we have seen in Section~\ref{sec:UV and 23S}, UV absorption by heavy ions is another crucial parameter that is difficult to give an accurate estimate. Observations that clarify the UV flux at $> 4.8\,{\rm eV}$ will be crucial for a better modeling of the He line feature at early times. Such spectra at early epochs are important also for Sr modeling as we discuss in \ref{sec:early}.

For Sr, the population ratio between $4{\rm D}$ and the ground levels follows the Boltzmann distribution with the photospheric temperature. Therefore, a higher temperature gives a higher abundance of the $4{\rm D}$ levels. The relevant transitions are 5S-5P and 4D-5P $(\rm 3.0\ and\ 1.2\,eV$: see Figure~\ref{fig:schematic_Sr}). The wavelengths are close to the optical wavelength, which is not affected by the UV cut-off in the spectrum. The observed flux of the photons that mediate these transitions is not much different from that of blackbody emission. Therefore, we expect that our modeling by blackbody flux sufficiently captures the physics.

\subsubsection{Photospheric density}
\label{sec:Photospheric density}
Ejecta density determines the recombination rate. Higher ejecta density gives a stronger spectral feature for both elements. The effect is crucial for Sr because higher density efficiently suppresses ionization. Sr models produce results consistent with the observation if we enhance the total Sr mass by a few to 10 times from 2.42 to 4.40 days, whereas the amount should be decreased by half for 1.43 days. We regard such an ad-hoc reduction/enhancement by a factor of 10 as unphysical because we have no other reason to assume these conditions. Still, some effects (e.g., ejecta stratification, peculiar density structure, opacity evolution, etc.) that our simple model does not capture may realize such density evolution.

\subsubsection{Nonthermal ionization rate}

We assume that the radioactive energy contributes to the ionization via beta particles and the radioactive heating rate is uniform across the ejecta. However, the nonthermal ionization rates become lower if most of it comes out as kinetic energies of heavier particles such as $\alpha$ particles or fission fragments \citep[see, e.g.,][]{2018Wanajo,WuPhRvL,Hotokezaka2020ApJ}. We also note that the radioactive heating rate around Sr, i.e., $80\lesssim A\lesssim 110$ is significantly weaker around $1$--$10$ day than that of $A\geq 110$. If elemental stratification exists in the ejecta and it spatially separates Sr from heavier elements, the ionization rate for Sr can be significantly lower.    
The low ionization rate allows more Sr to remain singly ionized, making absorption features stronger for Sr models at later times.

\subsubsection{Relativistic effect}
\label{sec:relativity}

The line formation region can be as fast as $\sim 0.3\,{\rm c}$. In such fast-moving ejecta, we need to consider the relativistic Doppler effect. The Doppler effect changes the effective temperature of the blackbody emission \citep{1986RybickiLightman}. The blackbody emission is isotropic on a frame comoving with the photosphere. The expansion velocity of absorption line forming region is always faster than the photosphere. Therefore, the relativistic effect would lower the temperature of the blackbody emission in the ejecta frame. It could work to decrease the UV flux in the first two epochs to help populate the $2^{3}{\rm S}$ level of the He atom.

\section{Conclusion}
We have shown that He atoms in kN can synthesize the P-Cygni profile observed in AT2017gfo if the mass fraction of He  is $X_{\rm He} \sim 0.2\,\%$. Note that a wide range of $X_{\rm He}$ is obtained in the merger outflow simulations in the literature. In particular, $X_{\rm He}\gtrsim 0.1\%$ is expected from the $\alpha$-rich freezeout of neutron-rich ejecta with an entropy of $\sim 20\,[k_B\,{\rm nuc}^{-1}]$. 
We model the spectra including the non-LTE effects, e.g., radioactive ionization by $\beta$ particles, which increases the ionization degree in the line forming region. The recombination of electrons and singly ionized He populates the metastable triplet level, which is responsible for the absorption line He I 10833${\rm \AA}$. Our model reproduces the line optical depth at the  epochs of $2.4$ and $3.4$ days. A caveat in the He interpretation for the observed line is that the line strength at the earlier epochs, e.g., 1.43 days, is rather sensitive to the photoionization rate of the metasable level. In fact, the He I is expected to be absent at 1.43 days if we assume that the continuum radiation has a black body spectrum with $4400\,{\rm K}$ and its Wien tail ionizes He I. However, the sharp cut-off in the observed spectrum around 4000\,${\rm \AA}$ at 1.43 days indicates that the photoioinzation rate is much lower than the expectation from the black body spectrum. Observations of the UV flux at early epochs will be beneficial for correctly modeling the photoionization, which will be useful for probing the mass ejection mechanism from ejecta entropy.

We have also explored the line formation of Sr II triplet around $10000\,{\rm \AA}$. 
Although the Sr II lines are expected be among the strongest lines in the LTE condition \citep{2019Watson,Domoto2022}, the line strength can be significantly reduced by radioactive ionization in the line-forming region. We found that the line signature is likely the strongest at $\sim 1.5$ days and gets weaker  with time because the efficiency of radioactive ionization increases with time. We showed that the Sr model can reproduce the observed spectrum at 1.43 days if the Sr mass is $\sim 1\%$ of the ejecta mass, which is roughly expected from the solar $r$-process pattern. However, we found that  the model fails to reproduce the spectra at the later epochs unless the mass fraction of Sr around the photosphere increases with time. Alternatively, it is possible that Sr II dominates the line feature at the earlier times while He I dominates at the later times. 
The two transitions of Sr\,II at $\sim 4000$\,\AA\ at early times will help infer the amount of Sr in the ejecta and its contribution to the P-Cygni profile we discuss. 
Optical properties of light {\it r}-process elements 
also help constrain the amount of Sr in kN ejecta.

\begin{acknowledgments}
We would like to thank Daiji Kato, Izumi Murakami, Sho Fujibayashi, Daniel Siegel and Kyohei Kawaguchi for fruitful discussion. We would like to thank Brian Metzger for his insightful comments on He production from high-entropy component of the ejecta. This research was supported by JST FOREST Program (Grant Number JPMJFR212Y, JPMJFR2136), NIFS Collaborative Research Program (NIFS22KIIF005), the JSPS Grant-in-Aid for Scientific Research (19H00694, 20H00158, 21H04997, 20K14513, 20H05639, 22JJ22810). N.D. acknowledges support from Graduate Program on Physics for the Universe (GP-PU) at Tohoku University.
\end{acknowledgments}

%

\vspace{5mm}





\appendix

\section{Physical processes}

\subsection{Atomic transitions}
\label{sec:atomic}
We assume that populating and depopulating processes are in equilibrium at each moment (steady state approximation): for each level $i$, we consider an equation $n_i \alpha_i = X_i$, where $X_i$ and $\alpha_i$ include all the populating and depopulating processes. They are electron collision excitation/de-excitation, photoionization, recombination, and ionization by non-thermal particles. Electron collision ionization is found to be negligible.

The transition rate by electron collision is computed as \citep{1987Berrington}:
\begin{equation}
q(i\rightarrow f) = \left\{
\begin{array}{ll}
\frac{8.63\times 10^{-6}}{w_i \sqrt{T}}\gamma_{if} e^{-(E_f-E_i)/kT} & (E_i < E_f)\\
\frac{8.63\times 10^{-6}}{w_i \sqrt{T}}\gamma_{if} & (E_i > E_f)
\end{array}
\right.
\end{equation}
where $w_i, \gamma_{if}$ are degrees of freedom for the state $i$ and the effective collision strength of order one.

Photoionization rate $R_{\rm photoion}$ from each level $i$ is computed as 
\begin{equation}
    R_{\rm photoion} = W\int^{\infty}_{E_{\rm th}}\sigma_{\rm photoion}(E_{\nu})\frac{B_{\nu}(E_{\nu}, T)}{E_\nu}dE_{\nu},
\end{equation}
where $W, \sigma_{\rm photoion}(E_{\nu}),$ and $B_{\nu}(E_{\nu}, T)/E_{\nu}$ are the geometric dilution factor, photoionization cross section from level $i$, and the number density of photons with energy $E_{\nu}$ in blackbody radiation.

Recombination rate $R_{\rm recombination}$ from ionization level $i+1$ to $i$ is computed as 
\begin{equation}
    R_{\rm recombination} = n_e n_{i+1} \alpha_i(T)
\end{equation}
where $\alpha_i(T)$ is the recombination coefficient \citep{1962Bates, 2010Nahar}.

For ionization rates by non-thermal particles $R_{\rm non-thermal}$, we first compute the heating rate per particle as $\Gamma = \dot{Q}/\mu m_{\rm H} [{\rm eV/s/particle}]$, where the mean atomic weight $\mu$ is assumed to be 100. We then compute the ionization rate of a particle $i$ as 
\begin{equation}
    R_{\rm non-thermal} = \frac{\Gamma}{w_{i}},
\end{equation}
where $w_i$ is the work per ion pair. 
For He, we use $w_{\rm HeI} = 593\,{\rm eV}$ and $w_{\rm HeII} = 3076\,{\rm eV}$. For Sr, the values are $w_{\rm SrI} = 124\,{\rm eV}$, $w_{\rm SrII} = 272\,{\rm eV}$, $w_{\rm SrIII} = 444\,{\rm eV}$, $w_{\rm SrIV} = 608\,{\rm eV}$, and $w_{\rm SrV} = 822\,{\rm eV}$ (see Hotokezaka et al. in prep. for derivation).

\subsection{Optical depth estimate}

\label{sec:optical depth from spectrum}

\begin{figure}
    \centering
    \includegraphics[width=0.5\columnwidth]{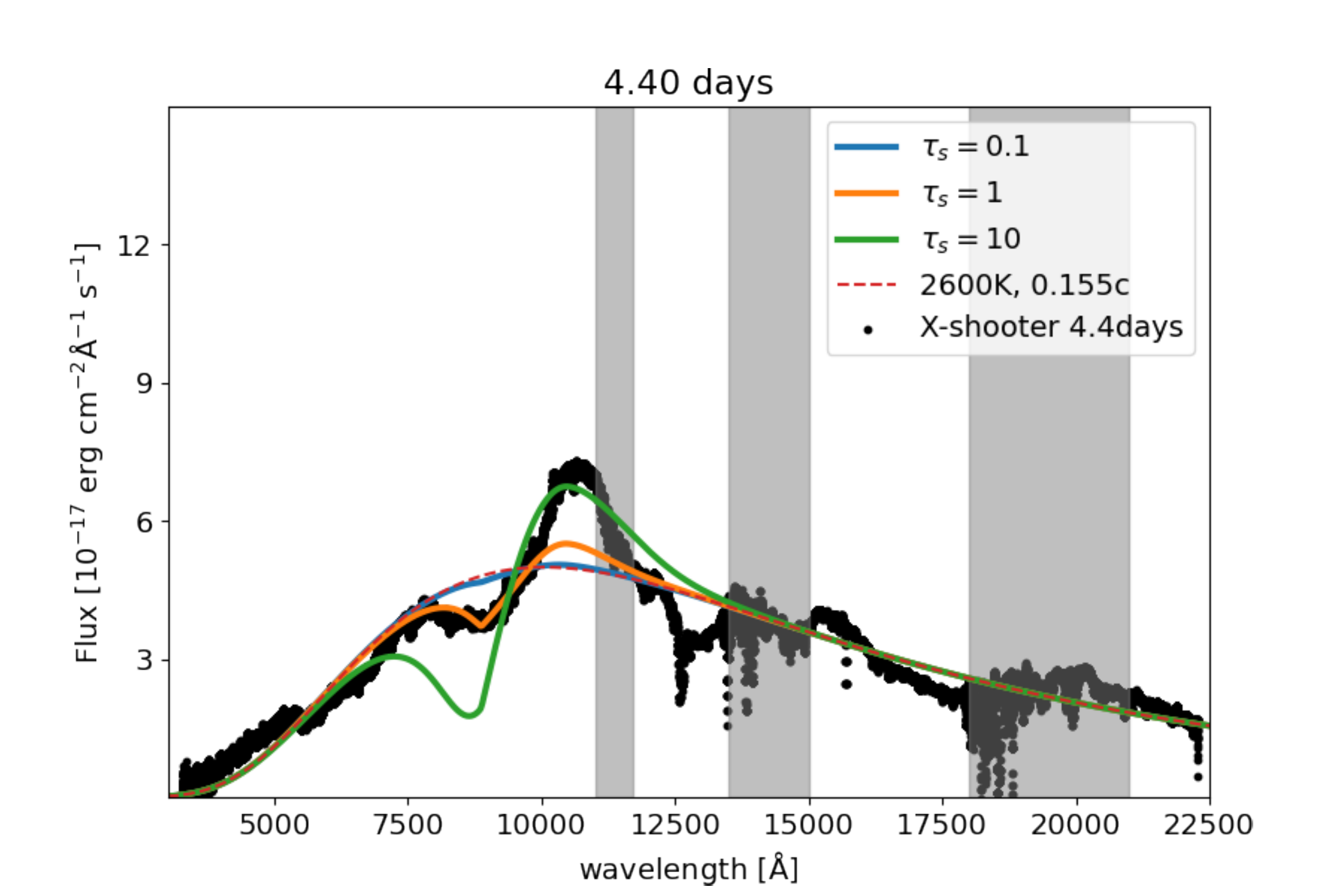}
    \caption{P-Cygni profile of the observed spectrum at 4.4 day after merger and the theoretical curves 
    with different values of the Sobolev optical depth at the photosphere, $\tau_s=0.1, 1, 10$. Here, we assume the photospheric emission as thermal radiation with a temperature of $2600\,{\rm K} $ and a photospheric velocity of $0.155c$. The gray shaded regions show the wavelength where telluric absorption is
    significant.}
    \label{fig:PCygni}
\end{figure}

Spectra of AT2017gfo show prominent P-Cygni profiles through 1.43 to 4.40 days. Figure~\ref{fig:PCygni} compares the spectrum of AT2017gfo at 4.40 days from the merger and the simplest model spectrum with various optical depths $\tau_s$. Here, we assume that the optical depth decreases as $\tau_s = \tau_{s0}(v/v_{\rm phot})^{-5}$, where $v_{\rm phot}$ is the photospheric velocity (see Section~\ref{sec:geometry}), and take $\tau_{s0}$ as a free parameter. The observation requires $\tau_{s0} \simeq 1$ at all the epochs between 1.43 to 4.40 days, no matter what element produces the feature. The value of $\tau_{s0}$ could be modified by a factor of $\sim 3$ by taking a different power-law index of the optical depth.

\subsection{Effect of UV blanketing for He}
\label{sec:UV and 23S}

\begin{figure}
    \centering
    \includegraphics[width=0.5\columnwidth
    ]{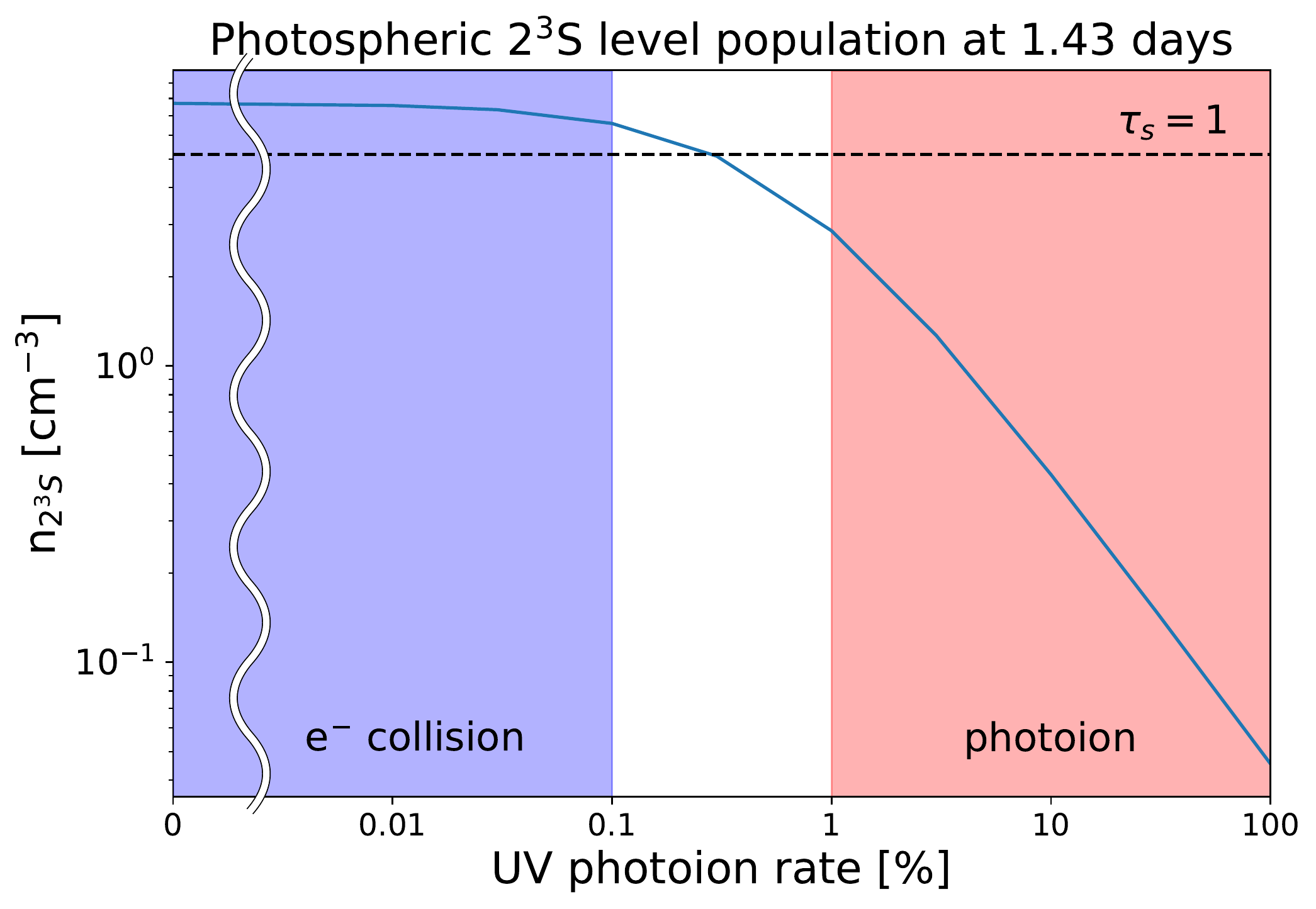}
    \caption{Photospheric density of He$_{\rm 2^{3}S}$ as a function of UV strength. Photoionization is dominant if UV strength is more than 1\% of blackbody emission, while electron collision is dominant if it is less than 0.1\%.}
    \label{fig:UV_23S}
\end{figure}

Figure~\ref{fig:UV_23S} shows the dependence of photospheric density of He\,I$_{2^{3}\rm S}$ level. The UV strength needs to be less than $\sim 0.3\%$ of the blackbody radiation to explain the observation at the first epoch. A similar estimate for the 2.42 days gives the maximum UV strength to be $\sim 30\%$. 


\bibliography{rprocess}{}
\bibliographystyle{aasjournal}



\end{document}